\documentclass[aps]{revtex4} 
\input{epsf}  

\usepackage{graphicx}

\def\vp{{{\bf \xi}}}
\def\mut{{\tilde{\mu}}}
\def\Ut{{\tilde{U}}}
\def\Vt{{\tilde{V}}}
\def\Wt{{\tilde{W}}}
\def\htt{{\tilde{h}}}
{\bf }
 
\newcommand{\be}{\begin{equation}}
\newcommand{\ee}{\end{equation}}
\newcommand{\bear}{\begin{eqnarray}}
\newcommand{\eear}{\end{eqnarray}}

\begin{document} 

\title{Crust-core coupling in rotating neutron stars}
  
\author{Kostas Glampedakis and Nils Andersson} 
\affiliation{School of Mathematics, University of
Southampton, Southampton SO17 1BJ, UK}

\date{\today}

\begin{abstract}
Motivated by their gravitational-wave driven instability, we investigate the 
influence of the crust on r-mode oscillations in a neutron star. 
Using a simplistic model of an elastic neutron star crust with constant shear modulus,
we carry out an analytic calculation with the main objective of deriving an
expression for the ``slippage'' between the core and the crust.      
Our analytic estimates support previous numerical results, and provide useful 
insights into the details of the problem.
\end{abstract}

\maketitle



\section{Introduction}

The study of neutron star oscillation modes has long been a vivid research area, attracting
additional attention in recent years after the discovery that inertial modes
of rotating fluid stars are prone to unstable behaviour when coupled to gravitational
radiation \cite{na98,morsink}. This instability is  characterised by growth timescales
short enough  to make it astrophysically relevant. For example, unstable  r-modes (the
class of purely axial inertial modes) have been proposed as a possible mechanism for limiting the
spin-periods of neutron stars in Low-Mass X-ray Binaries (LMXBs) and as a potential source of gravitational 
radiation for the ground-based detectors. For a detailed review of the r-mode instability we refer
the reader to Ref.~\cite{na_rev}.   

Studies of the r-mode instability have revealed a host of possible damping mechanisms 
counteracting the growth of an unstable mode. 
Among the most efficient is the damping due to friction, via the formation 
of an Ekman boundary layer at the crust-core interface, see \cite{Ekman} for a recent discussion. 
For mature neutron stars, this mechanism could be 
the key damping agent for unstable r-modes.  
The crust forms once the star cools below $10^{10}$~K or so,  soon after the star is born.
At the temperatures relevant for a mature neutron star, the heavy constituents in the core (neutrons, protons, 
hyperons) are expected to be in a superfluid/superconducting state. 
This leads to the suppression of, for example, the hyperon bulk viscosity \cite{haensel,reis,owen2} 
and the emergence of mutual friction (see also \cite{shear} for a recent discussion of the
shear viscosity in a superfluid neutron star core). In this paper we focus on the role of the crust. 
A detailed discussion of the natural first approximation, where the crust is
modelled as a rigid spherical ``container'', and the viscous Ekman layer that forms at the core-crust 
interface, can be found in \cite{Ekman}. 

In the present paper we explore a different aspect of the problem, namely, 
we consider a more refined model with an elastic crust characterised by a finite shear modulus. 
A model along these lines was considered by Lee \& Strohmayer \cite{LS} (hereafter LS) in their 
numerical study of oscillating neutron stars (for some earlier studies see \cite{mcdermott} 
and \cite{strohmayer}). Being elastic, the crust can sustain forced oscillations driven 
by (say) a core mode. In addition the crust has its own unique dynamics, with a set of oscillation modes
associated with the elasticity. The problem becomes very rich 
since one expects resonances/avoided crossings  to take place when a core-mode frequency 
approaches a crust-mode frequency \cite{yoshida}, for example as the star spins down. 
This picture is certainly much more complicated than one with 
the fluid simply ``rubbing'' against a rigid boundary wall. 

If we model the crust in a more sophisticated manner we should also expect the
relevant r-mode damping timescales to be altered. This was emphasised
by Levin \& Ushomirsky (hereafter LU) \cite{LU} who used
the formalism of LS but considered the simpler case of a uniform density,
slowly rotating spherical star, with a constant shear-modulus crust.
They argue that the change in the r-mode damping timescale, compared to the solid 
crust case, can be quantified in terms of the relative 
motion between the crust and the core. Introducing the dimensionless
``slippage'' $ {\cal S}_c = (v-v_0)/v_0 $, where $v$ is the core fluid velocity 
immediately beneath the crust and $v_0$ is the corresponding velocity in a fluid star, 
LU argue that an approximate revised damping timescale is  the standard Ekman result 
multiplied by the factor $1/{\cal S}_c^2$.  
LU attacked the problem by direct numerical integration of the crust-core equations, 
and showed that the slippage  is $ {\cal S}_c \approx 0.05-0.1$ in a typical case. 
Hence, they found that the r-mode damping timescale could be more than 100 
times slower than for a solid crust. For rotation rates such that the inertial 
r-mode frequencies are close to the various shear modes in the crust, the slippage 
factor can however become much larger (of order unity). These results were confirmed in a later 
study by Yoshida and Lee \cite{yoshida}.    

Our aim in this paper is to study the interaction between the oscillating fluid 
core and the crust using purely analytical tools. Our decision to pursue the problem 
analytically was driven by a  desire to obtain a better understanding of  the intricate coupling to the crust, and the way that 
the final result scales with the different parameters.  Although our chosen strategy 
leads to a significant amount of algebra, we believe that the end result
justifies the effort. Certainly, the analytical relations we derive illustrate the 
underlying physics much clearer than a direct numerical calculation would.  
To achieve our aims, we make use of boundary-layer theory \cite{Ekman} combined with a truncated expansion 
in spherical harmonics for the perturbed fluid variables. 
In order to keep the calculation  tractable, we consider a uniform density, slowly rotating Newtonian star 
with a constant shear modulus crust. We study in detail how the crust is driven by a single r-mode 
in the core, and then compute the slippage and the induced modifications to the r-mode. 
Our formulation of the problem is very similar to that of LU. The main difference is that we assume 
that a purely axial r-mode in the core acts as the driving force, while LU included ${\cal O}(\Omega^2)$
(polar) corrections to their core mode ($\Omega$ is the rotation rate). Our study is also limited
to a crust with a relatively small shear modulus (more like a jelly than a solid). 
In this sense our analysis is less general than that of LU \cite{LU}, but 
we show that realistic estimates of the shear modulus support our assumptions. 
Where both calculations are valid, our results are in good agreement with those of LU 
(they are not identical due to the slightly different formulations of the problem).  

The paper is organised as follows. In Section~\ref{sec:modulus} we briefly discuss the elastic
character of the crust,  providing some justification for our assumption of a constant
shear modulus. The main scope of the paper is developed in Section~\ref{sec:coupling} where we 
discuss neutron stars with elastic crusts, and study the delicate coupling between the core-flow and 
the crust. The concluding Sections, \ref{sec:results} \& \ref{sec:conclusions}, of the paper 
provide a discussion of the physical relevance of our results and a summary. In these sections we compare our 
results to previous work on this problem. We also point out interesting extensions of the present work. 

Aware of the somewhat heavy algebra involved in our analysis, we have made an effort to make the 
sections discussing the results self-contained. This should make the results accessible also for 
readers who are not prepared to bear with us as we work through the details of the core-crust problem. 
Many of the technical details of our calculation are located in the Appendices.


\section{Modelling the neutron star crust}
\label{sec:modulus}

We first discuss in some detail the properties of an elastic neutron star crust. This is the case 
that is relevant for most neutron stars since the crust is thought to form 
already at temperatures near $10^{10}$~K, possibly within minutes 
after the neutron star is born. Furthermore, the  crust is not expected 
to be particularly rigid. Even though it supports shear stresses its nature is more 
like jelly, and one would expect it to take part in any global oscillation of the star. 
In addition, it will support distinct crust modes characterised 
by the speed of shear waves in the nuclear lattice.  

\begin{figure}[ht]
\centerline{\includegraphics[height=6cm,clip] {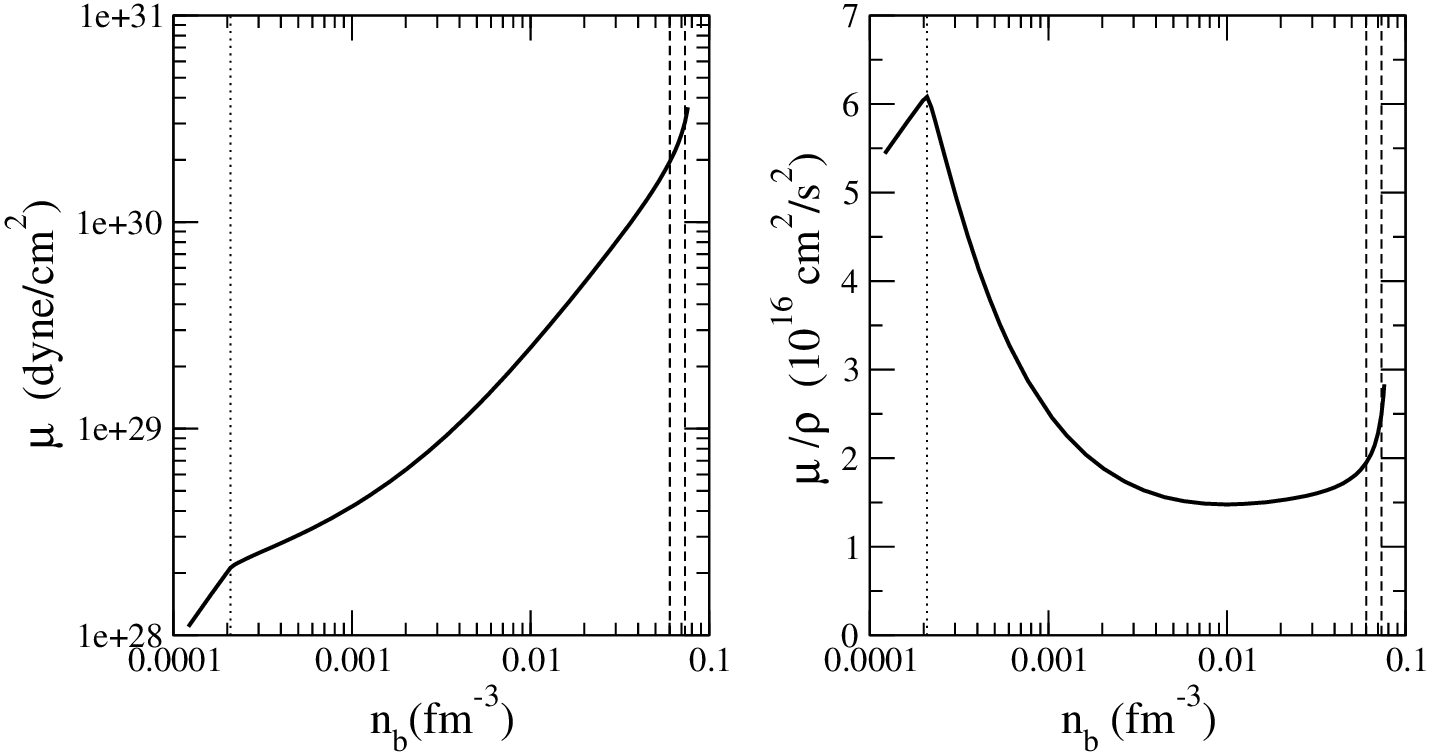} }
\caption{The shear modulus, based on Eq.~(\ref{shm}) and data for the 
equation of state of Douchin and Haensel \cite{douche}. The vertical 
dotted line indicates the location of neutron drip, while the dashed 
lines indicate the region near the base of the crust 
where the nuclei undergo a series of changes in shape, cf.
the discussion in \cite{douche}.}
\label{shearmod}
\end{figure}

In order to keep the problem analytically tractable, we will  adopt a simple 
uniform density neutron star model. In addition, we will take the fluid to be 
inviscid since we are mainly interested in the mechanical, dissipation free, 
coupling between the r-mode and the crust. Furthermore, we assume that the crust is 
characterised by a {\it uniform} shear modulus $\mu$. This assumption 
(which was also adopted by LU \cite{LU}) is somewhat unrealistic \cite{LS}, 
but it greatly simplifies the analysis. A study of the non-isotropic case requires a 
numerical solution. On the other hand, realistic models for the crust suggest that the speed of 
shear waves is remarkably constant throughout the crust. We illustrate this in 
Figure~\ref{shearmod} where we have combined the standard expression for the shear 
modulus of a {\it bcc} lattice
\be
\mu = 0.3711 { Z^2 e^2 n_N^{4/3} \over 2^{1/3}} 
\label{shm}\ee
where $Z$ is the number of protons per nucleus and $n_N$ is the number density of nuclei, with the 
data for the recent equation of state of Douchin and Haensel \cite{douche}.
As is clear from this figure the ratio
\be
\mut = { \mu \over \rho} \approx 10^{16} \mbox{ cm}^2/\mbox{s}^2
\ee
only varies by a factor of a few in the crust. It should also be noted that 
$\mu$ does not vanish at the core-crust 
phase-transition. Given the results shown in Figure~\ref{shearmod} 
it is clearly not too unreasonable to treat $\tilde{\mu}$ as a constant. 

The crust's rigidity can be quantified by means of the ratio
(essentially the ratio between the speed of shear waves and a characteristic
rotation velocity, or the relation between the elasticity and the Coriolis force),
\be
\Lambda \equiv \frac{\tilde{\mu}}{R^2 \Omega^2} 
\label{ratio} 
\ee
The crust behaves almost like a fluid when $ \Lambda \ll 1$ and is essentially
rigid in the opposite limit. In the present work we will only be concerned with the
former limit. In this sense, our analysis is less general than that of LU who put 
no restrictions on the ratio (\ref{ratio}). 
Introducing a characteristic rotation frequency 
\be
\Omega_0^2 = { G M \over R^3} \rightarrow \Omega_0^2 R^2 \approx  10^{20} 
\mbox{ cm}^2/\mbox{s}^2  
\ee
(such that the Kepler break up rotation rate is about $0.6 \Omega_0$)
we see that we will have 
\be
\Lambda \approx 10^{-4} \left( { \Omega_0 \over \Omega} \right)^2
\ee
This clearly shows that the $\Lambda \ll 1$ limit is appropriate provided that 
we do not let $\Omega \to 0$. In a sense, this  illustrates the fact that 
our calculation is limited to rotating stars. Another reason for this is that 
we will assume that the core fluid undergoes an inertial r-mode oscillation. 
This would not make sense for a non-rotating star as these modes then become trivial.


\section{Crust-core coupling}
\label{sec:coupling}

\subsection{Formulating the problem}

The physical situation we wish to examine is that of a single core r-mode
coupled to the crust. As a result of this coupling the crust will 
oscillate. At the same time there will be some ``back-reaction'' on the 
core mode. Mathematically, one needs to solve the  Euler equation in
the core region, and the corresponding equations for an elastic body 
in the crust region.  
Boundary conditions need to be imposed at the core-crust interface 
and at the star's surface and centre. Our aim is to approach the problem 
through a two-parameter expansion: a slow rotation expansion in 
$\Omega/\Omega_0$ together with an expansion in $\Lambda$. The discussion in
the previous Section illustrates the reason why this is a tricky problem. 
Yet we will see that we can make significant (possibly surprising) progress by combining 
these expansions with boundary layer techniques.

We assume a uniform density, incompressible fluid. Since we work at leading order
in rotation  we can assume our star to be spherical. Moreover, we adopt the
so-called Cowling approximation where the perturbation in the gravitational potential
is ignored. This is a sensible assumption, since for the quasi-axial modes we are 
interested in the associated density perturbation (and consequently the perturbation in the
gravitational potential) is a higher order correction. In the rotating frame, the linearised Euler 
and continuity equations for the crust flow are
\bear
&& \partial_t^2 {\bf \xi} + 2 {\bf\Omega} \times \partial_t {\bf \xi} =  
- { 1 \over \rho} \nabla \delta p + \tilde{\mu} \nabla^2 {\bf \xi}
\label{navier}
\\
&& \nabla \cdot{\bf \xi} = 0
\label{conti}
\eear
where we have taken $\tilde{\mu}= \mu/\rho=$~constant. 
For the present problem it is natural to work with the displacement vector $\vp$. The main reason 
for this is that the shear introduces a Hookean restoring force into the 
Euler equations, i.e. terms $\sim \mu \nabla^2 \xi$ \cite{LS,strohmayer,mcdermott}.
The core flow is described by
the $\mu \to 0 $ limit of the above equations.

We assume a decomposition of the form,
\begin{equation}
{\bf \vp} = \frac{1}{r} \sum_{l \geq m} \left [\, W_l Y_l^m {\bf \hat{e}_r}  + \left( V_l \partial_\theta Y_l^m
-  i{U_l \partial_\varphi Y_l^m \over \sin \theta} 
\right) {\bf \hat{e}_\theta}
+   \left( { V_l \partial_\varphi Y_l^m \over \sin \theta}  
+ i U_l\partial_\theta Y_l^m \right) 
{\bf \hat{e}_\varphi} \right ] e^{i\sigma t}
\label{decomp}
\end{equation}
where the relative phase between the polar $\{W_l, V_l \} $ and axial components $\{U_l \} $ 
is chosen such that the final perturbation equations are real valued.
We also decompose the perturbed pressure as 
\begin{equation}
\frac{\delta p}{\rho} = \sum_{l \geq m} h_l Y_l^m
\end{equation}

The three components of the Euler equation (\ref{navier}) then reduce to
(in the following, a prime denotes a radial derivative);
\begin{eqnarray}
{\cal E}^{r} &=& \sum_{l \geq m} \left [ \left \{  r h_l^\prime - \sigma^2 W_l  
+2m\sigma\Omega V_l 
\right\} Y_l^m + 2\sigma\Omega  U_l  \sin \theta \partial_\theta Y_l^m 
+ \tilde{\mu} \left\{ 2{W_l \over r^2} -  W_l^{\prime\prime}  + 
l(l+1) {W_l - 2V_l \over r^2} \right\}Y_l^m \right ]   = 0
\label{r_Euler}
\nonumber \\ 
\\
\nonumber \\
{\cal E}^\theta &=& \sum_{l \geq m} \left [ -m \sigma^2  U_l  Y_l^m 
+ \left\{ h_l - \sigma^2  V_l \right\} \sin \theta \partial_\theta Y_l^m +
2 m  \sigma \Omega V_l  \cos \theta Y_l^m + 2\sigma \Omega  U_l \cos \theta \sin \theta \partial_\theta Y_l^m  
\right.
\nonumber \\
&&- \left. m \tilde{\mu} \left\{ U_l^{\prime\prime}  
- l(l+1) { U_l \over r^2} \right\} Y_l^m - 
\tilde{\mu} \left\{ V_l^{\prime \prime}  - l(l+1) { V_l \over r^2} + 
2 { W_l \over r^2} \right\} 
\sin \theta \partial_\theta Y_l^m \right ]
= 0
\label{th_Euler}
\\
\nonumber \\
\nonumber \\
{\cal E}^\varphi &=& \sum_{l \geq m} \left [  \left\{ m h_l  + 2 \sigma \Omega  W_l 
- m \sigma^2  V_l     \right\} Y_l^m - \sigma^2 U_l \sin \theta \partial_\theta Y_l^m
 + 2m\sigma \Omega  U_l \cos \theta Y_l^m  + 2 \sigma \Omega  V_l \cos \theta \sin \theta 
\partial_\theta Y_l^m \right. 
\nonumber \\
&&- \left.  2 \sigma \Omega  W_l  \cos^2 \theta  Y_l^m 
-m \tilde{\mu} \left\{  V_l^{\prime \prime}  - 
l(l+1){ V_l \over r^2}  + 2{ W_l \over r^2}\right\} Y_l^m
- \tilde{\mu} \left\{ U_l^{\prime \prime}  - 
l(l+1){ U_l \over r^2} \right\} \sin \theta \partial_\theta Y_l^m \right ]
= 0
\label{ph_Euler}
\end{eqnarray}
 These equations are identical to those given in 
\cite{LS,strohmayer}, specialised for constant shear modulus and incompressible flow.

As discussed by Lee \& Strohmayer \cite{LS} it is convenient to consider two 
particular combinations of the above equations. The first is the radial component 
of the vorticity,
\be
m {\cal E}^\theta - \sin \theta \partial_\theta {\cal E}^\varphi =0
\label{combo1} 
\ee
and the second is the ``divergence equation'',
\be
\sin \theta \partial_\theta {\cal E}^\theta - m {\cal E}^\varphi = 0 
\label{combo2}
\ee
From eqn.~(\ref{combo1}) and from the combination of eqns.~(\ref{combo2}) and (\ref{r_Euler})
we arrive at the following equations,
\bear
&& \frac{\sigma}{2\Omega} \left[ l(l+1) \frac{\sigma}{2\Omega} - m \right] U_l 
+ \frac{\sigma}{2 \Omega} \left [\, (l+1) Q_l \{ W_{l-1} - (l-1) V_{l-1} \}   
-  l Q_{l+1} \{ W_{l+1} + (l+2) V_{l+1} \}\, \right ] 
\nonumber \\
&& + l(l+1) \frac{\tilde{\mu}}{4\Omega^2} \left[  U_l^{\prime\prime} - l(l+1) { U_l \over r^2} \right] = 0
\label{eq1}
\\
\nonumber \\
\nonumber \\
&& \frac{\sigma}{2\Omega} \left [ l(l+1) \{ (l-1) Q_l U_{l-1} - (l+2) Q_{l+1} U_{l+1} \} 
-(l-1)(l+1) Q_l r U_{l-1}^\prime - l(l+2) Q_{l+1} r U_{l+1}^\prime 
-m \{ r V_l^\prime - W_l \}  \right ]
\nonumber \\
&& +  l(l+1) \left ( \frac{\sigma}{2\Omega} \right )^2 \left [ rV_l^\prime - W_l \right ] 
+ l(l+1) \frac{\tilde{\mu}}{4\Omega^2} \left [ r V_l^{\prime\prime\prime} 
- 2W_l^{\prime\prime} + (l-1)(l+2){W_l \over r^2} \right ] = 0
\label{eq2}
\eear
where
\be
Q_l = \sqrt{\frac{(l+m)(l-m)}{(2l-1)(2l+1)}}
\ee 
 
The continuity equation (\ref{conti}) provides one additional expression
\be
(rW_l)^\prime - l(l+1) V_l = 0
\label{continuity}
\ee

Equations (\ref{eq1}), (\ref{eq2}) and (\ref{continuity}) are the basic equations for our 
problem. They involve only the eigenfunctions of ${\bf \xi} $ and the frequency $\sigma $ and are 
equivalent to the initial Euler equations. It is interesting to note that our equations only couple 
multipoles of order $l$ to those corresponding to $l\pm1$. This is in contrast to the (fluid) equations used 
by eg. Lockitch and Friedman \cite{lockitch} to study inertial modes, where coupling to $l\pm 2$ multipoles 
was also present. In this sense, the present set of equations is simpler and could possibly be advantageous also 
in studies of the fluid inertial mode problem.  


\subsection{Boundary conditions}
\label{sec:b_conditions}

The  dynamical equations are accompanied by boundary conditions at the crust-core interface (at $r=R_c$) and at 
the surface of the star ($r=R$). These boundary conditions can be formulated in terms of the traction 
components \cite{landau,LS}, 
\bear
T_r &=& 2\mu \partial_r \xi^r -\Delta p 
\\
\nonumber \\
T_{\theta} &=& \mu \left [ \partial_r \xi^{\theta} -\frac{1}{r} \xi^{\theta}
+ \frac{1}{r} \partial_{\theta} \xi^r  \right ]
\\
\nonumber \\
T_{\varphi} &=& \mu \left [  \partial_r \xi^{\varphi} -\frac{1}{r} \xi^{\varphi}
+ \frac{1}{r\sin\theta} \partial_{\varphi} \xi^r  \right ]
\eear 
where $\Delta p$ is the Lagrangian variation in the pressure. The traction vector is constructed by 
projecting the fluid stress tensor along the unit normal vector at the interface, and represents the force 
per unit area. Perturbing (in the Lagrangian sense) this projection results in the expressions above.  

At the crust-core interface all traction components as well as the radial displacement vector 
need to be continuous. Specifically, this requirement leads to (after setting $l=m $ in the axial component)\\
\noindent
(i) Vanishing tangential traction at the interface and surface:
\be
\mut [ r U_m^\prime -2U_m ] =0 \qquad \mbox{at} \quad r= R_c,~ R 
\label{bc1}
\ee 
\be
\mut \left [ r V_{m+1}^\prime + W_{m+1} -2 V_{m+1}  \right ] =0 \qquad \mbox{at} \quad 
r=R_c, ~ R  
\label{bc2}
\ee 
(ii) Continuity of the radial traction at the interface:
\be
\delta p^{\rm core} = \delta p^{\rm crust} -2 \mu \partial_r \xi^r \quad \Rightarrow \quad
h_{m+1}^{\rm core} = h_{m+1}^{\rm crust} -2\mut \left ( \frac{W_{m+1}}{r} 
\right )^\prime \qquad \mbox{at} \quad r=R_c 
\label{bc3}
\ee
(iii)  Vanishing of the radial traction at the surface:
\be
\delta p + \xi^r \nabla_r p -2\mu \partial_r \xi^r =0  \quad \Rightarrow \quad
R^2 h_{m+1} = -g_0 R W_{m+1} + 2\mut [ R W_{m+1}^\prime - W_{m+1} ] \qquad \mbox{at} \quad r= R 
\label{bc4}
\ee
where we have defined  $ g_0 = GM/R^2 $, the surface gravitational acceleration
(which appears after using the background Euler equation 
${\bf \nabla} p = - \rho {\bf \nabla} \Phi $). One can show that this final condition 
reduces to the standard condition of vanishing Lagrangian pressure perturbation 
at the surface in the limit $\mut \to 0$.\\
(iv) Continuity of the radial displacement at the interface:
\be
W_{m+1}(R_c^{-}) = W_{m+1}(R_c^{+})
\label{bc5}
\ee
In addition, we require all core flow variables to be finite at
$r=0$. 


\subsection{Simple mode solutions?}

Before we begin our main analysis, it is worth discussing the nature of 
some comparatively simple solutions. First of all, we ought to be able to reproduce 
the standard fluid r-mode solution from equations (\ref{eq1}) and (\ref{eq2}).
Let $\tilde{\mu} = 0$ and assume that the 
only non-vanishing displacement is $U_m$. Then eqn.~(\ref{eq1}) with $l=m$ leads to 
\be
\left[ m(m+1) \sigma^2 - 2m\sigma \Omega \right] U_m = 0
\label{req1}\ee
and we must clearly have either $\sigma = 0$ or
\be
\sigma = { 2 \Omega \over m+1} 
\label{rfreq}\ee
which is the standard result. Meanwhile eqn.~(\ref{eq2}) with $l=m-1$ is trivially 
satisfied since $Q_m=0$. Finally we need eqn~(\ref{eq2})  with $l=m+1$;
\be
r U_m^\prime - (m+1) U_m   = 0 \quad \Rightarrow \quad U_m =  A r^{m+1}
\label{req2}\ee
which is the expected eigenfunction.

Given the perturbation equations from the previous section
it is also easy to prove that inertial modes 
in uniform density stars cannot have polar terms as their highest 
multipole contribution \cite{lockitch}. Consider a mode which truncates after 
$l=n$ (say) and for which the $n^\mathrm{th}$ terms are polar. Then eqn.~(\ref{eq1}) 
with $l=n+1$ provides the relation
\be
W_n-n V_n = 0
\ee            
Combining this with the continuity equation we see that we must have
\begin{equation}
W_n = r^n
\end{equation}
which cannot satisfy the required surface boundary condition $W_n(R)=0$.
In other words, the highest multipole term of an inertial-mode solution
cannot be polar. 

More important for the present investigation is to 
ask whether we can have purely axial modes in a star with a fluid core 
and an elastic crust. We have already shown that in order to have only $U_m$ (say) nonzero 
in the fluid core, we must have a mode with the r-mode frequency.  
We also have the eigenfunction from 
Eq.~(\ref{eq2}). In order to be consistent, this eigenfunction must satisfy
the remaining part of Eq.~(\ref{eq1}) in the crust, i.e.
\begin{equation}
U_m^{\prime\prime} - m(m+1) { U_m \over r^2} = 0
\end{equation}
This is, in fact, the case. But our solution is nevertheless not acceptable 
because it cannot satisfy the boundary conditions of vanishing traction at 
the base of the crust and the surface of the star [Eq.~(\ref{bc1}) 
above]. We require
\begin{equation}
U_m^\prime - 2 { U_m \over r} =0 \quad \mbox{ at } \quad r=R_c \quad \mbox{ and } 
\quad r=R
\end{equation}
which clearly is incompatible with our solution. Thus we conclude that {\it a 
purely axial mode is no longer possible in the presence of the crust}. 

This is, in fact, a bit puzzling because we can extend the argument to 
demonstrate that a crust mode with only $[U_m,V_{m+1},W_{m+1}]\neq0$ is
also inconsistent! This would seem to cast some doubt on the numerical results of LU
\cite{LU} who only include these terms. In reality the situation is a bit 
more complex. As also mentioned by Lee and Strohmayer \cite{LS}, the problem should 
be approached in a way similar to the inertial-mode problem for a fluid star \cite{lockitch}. 
Even though one should not at the outset restrict the number of  multipoles included 
in the calculation  one will always in practise truncate after some 
multipole. Since this involves neglecting some of the relations implied by 
the recurrence relations it renders the solution somewhat inconsistent
(as in the case of LU). However, one would expect the mode-solution to converge
when more multipoles are accounted for, and it does not seem unreasonable 
to expect that a lower order representation will be fairly accurate. Indeed,
this turns out to be the case in a similar calculation for the standard Ekman
layer problem where the known exact solution is quite well approximated
by a $[U_m,V_{m+1},W_{m+1}]$ truncated representation (see \cite{Ekman} for discussion). 
As we are seeking an analytical approximation to the solution, it is natural to truncate
the solution at the simplest "meaningful" level. This is an obvious weakness of our analysis. 
A more detailed study, investigating for example the convergence as more multipoles are accounted for
and whether the outcome depends on the parameter values, 
requires a numerical solution. 


\subsection{Core flow}
\label{sec:coreflow}

As we have mentioned, we will assume that the bulk core flow is described by a single $l=m$ r-mode, which 
is capable of penetrating the crust region and driving the flow there. In return, it picks up corrections 
due to the coupling with the crust. Furthermore, we will truncate the multipolar expansion 
(see eqn.~(\ref{decomp})) at the level $\{U_m, V_{m+1}, W_{m+1}\}$. Note that the same truncation
was adopted in previous studies of the problem  \cite{LS,LU}.

For the dynamical variables we assume expansions of the form 
\be
Q = Q^0 + Q^1 + Q^2 + ...
\label{core_expan}
\ee
where, for brevity, we have collectively denoted $Q= \{U_m,V_{m+1},W_{m+1},h_{m+1},\sigma \}$. 
Expansion (\ref{core_expan}) has no obvious small parameter in it; it is written anticipating  the
fact that there is a one to one correspondence with the similar expansion 
in the crust region (which has the well defined expansion parameter $\mut$, see Section~\ref{sec:crustflow}). 
The reason for including terms up to second order will  become apparent in the discussion of the crust flow. 

As we are assuming an r-mode solution at leading order we take the leading order polar pieces to be zero, i.e.
\be
 W^{0}_{m+1}=V^{0}_{m+1}=0
\ee
The non-zero axial piece is normalised in such a way that $A=1$ in (\ref{req2}), i.e. 
$ U^0_m = R_c^{m+1} $ at the base of the crust. 

The next step is to insert the expansion (\ref{core_expan}) in the main equations (\ref{eq1})-(\ref{continuity}).
The resulting first and second order equations are listed in Appendix~\ref{app:core}. The analysis in that Appendix
provides the results,
\bear
V^{1}_{m+1} &=& \frac{\sigma_1}{2\Omega} (m+1) Q_{m+1}  r^{m+1}
\\
\nonumber \\
W^{1}_{m+1} &=& (m+1) V^{1}_{m+1}
\\
\nonumber \\
U_{m}^{1} &=& \gamma r^{m+1}
\eear 
and
\bear
V_{m+1}^2 &=& \frac{[ \sigma_2 +\gamma\sigma_1]}{2\Omega} (m+1)Q_{m+1}  r^{m+1}
\\
\nonumber \\
W^{2}_{m+1} &=& (m+1) V^{2}_{m+1}
\\
\nonumber \\
U_m^2 &=& \delta r^{m+1}
\eear 
where $\gamma$ and $\delta $ are constants. 

These results demonstrate the importance of the frequency corrections $\sigma_1$ and $\sigma_2$. Had we 
neglected them, we would have effectively killed all corrections to the core flow. 
Hence, the secondary core flow is linked to the crust-induced correction to the mode oscillation frequency.

Finally, we need to calculate the pressure correction, anticipating the fact
that the pressure appears explicitly  in some of the boundary conditions. 
For this purpose it is convenient to use the ${\cal E}^\theta$ component of the 
Euler equation, eqn.~(\ref{th_Euler}). We easily obtain,
\bear
h_{m+1}^{0} &=& -\frac{m Q_{m+1}}{(m+1)^2}r^{m+1}
\label{pres1}
\\
\nonumber \\
h_{m+1}^1 &=& -\frac{4\Omega^2 Q_{m+1}}{(m+1)^2} \left [\, \frac{\sigma_1}{2\Omega} (m+1)(2m-1)
+ m \gamma\, \right ] r^{m+1}
\\
\nonumber \\
h_{m+1}^2 &=& -\frac{4\Omega^2 Q_{m+1}}{(m+1)^2} \left [\, (m+1)(2m-1)\frac{\sigma_2 +\gamma \sigma_1}{2\Omega} 
+ m \delta
+ (m-2)(m+1)^2 \left (\frac{\sigma_1}{2\Omega} \right )^2 \, \right ] r^{m+1}
\eear
In the end, all core-flow eigenfunctions (up to second order) are expressed in 
terms of the (as yet unknown) parameters $ \sigma_1 $, $\sigma_2$, $ \gamma$ and 
$\delta$. As we will see, these are linked to the crust flow through the interface 
conditions.


\subsection{Crust flow: setting up the equations}
\label{sec:crustflow}

In the crust region the Euler equations have additional term of form $ \mut \nabla^2 {\bf \xi}$. 
These terms are similar to the viscous terms in the Ekman layer problem \cite{Ekman}, but here they represent the 
crust's elastic restoring force. Despite the additional terms being dissipative in 
one case and oscillatory in the other, the problems are mathematically similar. 
The crust equations are likely to exhibit ``boundary layer behaviour'' when the 
coefficient $\mut$, which multiplies the highest order spatial derivatives,
is treated as a small parameter (i.e. in the limit $\Lambda \ll 1$).  
This feature will turn out to be of key importance when solving the  equations. 

Since we are only considering an ``almost fluid'' crust, one would expect the crust flow to  
be primarily axial in nature. as it is driven by the quasi-axial core oscillation. 
In the crust, as in the core, we assume from the outset that $V_{m+1}$ and $W_{m+1}$ vanish at leading order,
and truncate the multipole expansion (\ref{decomp}) by including only the $\{U_m,V_{m+1},W_{m+1}\} $ 
eigenfunctions.

In the crust, the continuity equation (\ref{continuity}) with $l=m+1$, 
gives
\be
 (r W_{m+1} )^\prime = (m+1)(m+2) V_{m+1}
\label{cont_crust}
\ee
Meanwhile (\ref{eq1}) with $l=m$, and (\ref{eq2}) with $l=m+1$,  can be written
\bear
&&\frac{\sigma}{2\Omega} \left [ (m+1) \frac{\sigma}{2\Omega} -1 \right ] U_m + \frac{\mut}{4\Omega^2} (m+1) \
\left [ U_m^{\prime\prime} -m(m+1) \frac{U_m}{r^2}  \right ] -\frac{\sigma}{2\Omega} Q_{m+1} [ W_{m+1} + 
(m+2) V_{m+1} ] = 0
\label{vort_crust}
\\
\nonumber \\ 
\nonumber \\
&& \frac{\sigma}{2\Omega} m \left [ (m+1)(m+2) Q_{m+1} U_m -(m+2) Q_{m+1} r U^\prime_m
-r V_{m+1}^\prime + W_{m+1}  \right ] 
\nonumber \\
&&  + \left ( \frac{\sigma}{2\Omega} \right )^2 (m+1)(m+2) \left [ rV_{m+1}^\prime  -W_{m+1} \right ] 
+ \frac{\mut}{4\Omega^2} (m+1)(m+2) \left [ r V_{m+1}^{'''} -2W_{m+1}^{''} + m(m+3) \frac{W_{m+1}}{r^2}  
\right ]
\label{div_crust}
\eear
We now adopt a standard uniform expansion \cite{orszag},
\be
Q = Q^0 + \tilde{Q}^0 + \mut^a \{ Q^1 + \tilde{Q}^1  \}  
+ \mut^{2a} \{ Q^2 + \tilde{Q}^2  \} + {\cal O}(\mut^{3a})
\label{uniform}
\ee
where we have again used the collective notation $Q=\{ U_m,V_{m+1},W_{m+1},h_{m+1}, \sigma\}$. 
``Boundary layer'' variables are denoted by tildes (and for them we assume dominant radial derivatives 
$ \partial_r \sim {\cal O}(\mut^{-a}) $). The remaining quantities pertain to the smoothly varying, 
background crust flow.  Note that in (\ref{uniform}) the power-law exponent $a$ is initially undetermined.
Its value is fixed by the equations themselves, in order for them to make sense.
The analysis of Appendix~\ref{app:crust} shows that $a = 1/2 $.  

Inserting the expansion (\ref{uniform}) in eqns.~(\ref{cont_crust})--(\ref{div_crust}) 
we find two families of equations, one for the ``boundary layer'' and one for background variables.
We derive  all the dynamical equations required for the determination of the crust flow up 
to ${\cal O}(\mut)$ (the reason for pursuing the calculation to this order, rather than simply
to ${\cal O}(\mut^{1/2})$ will become apparent below). Next, we expand the various 
boundary conditions (\ref{bc1})--(\ref{bc5}).  In order to avoid overloading 
the reader with algebra we carry out these calculations in 
Appendices~\ref{app:crust} and \ref{app:bcs}.


\subsubsection{Leading-order flow}

Having written down all the relevant equations, we can now solve eqns.~(\ref{vort1}) and 
(\ref{div1}) for the leading order flow. After decoupling these equations we end up with,
\be
\mut^2 \Vt_{m+1}^{0 \prime\prime\prime\prime} + \frac{8\Omega^2 \mut}{(m+1)^2 (m+2)} 
\Vt_{m+1}^{0 \prime\prime}
- \frac{16 \Omega^4 m(m+2)}{(m+1)^4} Q_{m+1}^2 \Vt_{m+1}^0 =0
\label{veq}\ee
Trying the Ansatz, $\Vt_{m+1}^0 \sim e^{\lambda (r-R_c)} $ we find for the 
exponent a pair of real and a pair of imaginary roots,

\bear
\lambda_{1,2} &=& \pm \frac{2\Omega}{\mut^{1/2}} \left( \frac{{\sqrt{\alpha} -1}}
{ (m+2) (m+1)^2} \right)^{1/2}
\label{root1}
\\
\nonumber \\ 
i \lambda_{3,4} &=& \pm \frac{2 \Omega}{\mut^{1/2}}\left( \frac{{\sqrt{\alpha}  + 1}}
{ (m+2) (m+1)^2} \right)^{1/2}
\label{root2}
\eear
where 
\be
\alpha = 1 + m(m+2)^3 Q_{m+1}^2 
\ee 

The occurrence of purely real and  imaginary roots deserves further
discussion. The former will lead to a pair of growing/decaying exponentials
while the latter implies a pair of oscillatory exponentials. This is not the typical 
situation in boundary layer problems. Recall, for example, that in the Ekman
problem \cite{Ekman} we have only decaying and growing exponentials.
The difference in the present problem is associated with the fact that the $\mut \nabla^2 {\bf \xi}$ term 
in the Euler equation represents an elastic restoring force, not a viscous damping force. 
Intuitively, the purely oscillatory solutions to (\ref{veq}) represent
the elastic degree of freedom of the crust, eg. allows for the presence of the
crustal shear modes in the calculation. This means that in order to properly construct the complete
 flow we need to keep all four solutions. Therefore we write, 
\be
\Vt_{m+1}^{0} = A_v e^{\lambda_1 \zeta} + B_v e^{-\lambda_1 \zeta}
+ C_v \cos(\lambda_3 \zeta) - D_v \sin(\lambda_3 \zeta)
\ee
where $\zeta = r-R_c$.   Similarly,
\be
\Ut_{m}^{0} = A_u e^{\lambda_1 \zeta} + B_u e^{-\lambda_1 \zeta}
+ C_u \cos(\lambda_3 \zeta) -D_u \sin(\lambda_3 \zeta)
\ee
As a result of eqns.~(\ref{vort1}) and (\ref{div1}), the coefficients  are coupled 
in the following manner, 
\bear
\mut  \lambda_1^2 (m+1)^2 \left\{ \begin{array}{ll} A_u \\ B_u \end{array} \right\} 
&=& 4\Omega^2 (m+2) Q_{m+1} \left\{ \begin{array}{ll} A_v \\ B_v \end{array} \right\}
\label{coeff1} \\
-\mut \lambda_3^2 (m+1)^2 \left\{ \begin{array}{ll}  C_u \\  D_u \end{array} \right\}
&=& 4\Omega^2 (m+2) Q_{m+1} \left\{ \begin{array}{ll} C_v \\  D_v \end{array} \right\}
\label{coeff3}
\eear
We can now impose the boundary conditions (\ref{axbc1}) and (\ref{pobc1})
at $r=R_c$. These translate into
\bear
\lambda_1 ( A_v - B_v) &=& \lambda_3 D_v
\\
\lambda_1 ( A_u -B_u ) &=& \lambda_3 D_u
\eear
It is easy to see that these two relations are incompatible with
(\ref{coeff1})  and (\ref{coeff3}). The only remaining 
option is the trivial result $D_u = D_v =0$, 
which in turn means that
\be
A_u = B_u , \qquad \mbox{and} \qquad A_v = B_v 
\ee
Enforcing  the conditions (\ref{axbc1}) and (\ref{pobc1}) again, 
this time at $r=R$, leads to
\bear
2\lambda_1 A_u \sinh(\lambda_1 \Delta) &=& \lambda_3 C_u\sin(\lambda_3 \Delta)
\\
2\lambda_1 A_v \sinh(\lambda_1 \Delta) &=& \lambda_3 C_v \sin(\lambda_3 \Delta)
\eear
where $ \Delta \equiv R-R_c$ is the crust thickness. Combining these relations with the previous 
ones we are led to the trivial result,
\be
C_v = C_u = A_u = A_v = B_u = B_v = 0 
\ee
However, the argument that non-trivial solutions do not exist is only true provided that 
\be
\sin(\lambda_3 \Delta ) \neq 0 \quad  \Rightarrow \quad \lambda_3 
\Delta \neq n \pi, \quad n~\mbox{integer}
\label{reson}
\ee
This condition is obviously  violated when $\Omega \rightarrow 0$. However, as we
have already emphasised, we are not allowed to approach
the nonrotating limit  in the present framework (since for a given $\mut$ we always 
demand that $\Lambda \ll 1$).
On the other hand, non-zero values of $\Omega$ that violate (\ref{reson}) could
well be accessible. Later on in this paper we discuss the physical meaning of these
particular  frequencies.   

We conclude that when (\ref{reson}) holds we must have
\be
\Ut_{m}^{0} = \Vt_{m+1}^{0}=0
\ee
It also follows that  $\htt_{m+1}^0=0$. In fact, by using eqns.~(\ref{vort1}) and
(\ref{div1}) in (\ref{pres2}), we can prove that $ \htt_{m+1}^0 =0 $ identically.
Hence, somewhat unexpectedly, we conclude that at the anticipated leading order all 
eigenfunctions must vanish.

Before we move on, we should perhaps emphasise why we think it was worth working
through the above calculation even though the end result was trivial. 
Basically, we wanted to highlight the \underline{exception} (\ref{reson}), i.e., 
the existence of non-trivial solutions for certain parameter values. 
It is also worth pointing out that we would not necessarily have expected  
that the solution would be trivial at this level of approximation. 
Having said this, the result could have been inferred from the  fact that the corresponding boundary 
conditions involve only ``boundary layer'' quantities \cite{orszag}.


\subsubsection{First order crust flow}

Having obtained a trivial result for the leading order flow, 
we must proceed to the next level (i.e. $ {\cal O}(\mut^{1/2})$). Let us first focus on the ``boundary layer'' 
variables. The continuity equation (\ref{cont2}) gives,
\be
\Wt_{m+1}^1 = 0
\ee
We can also calculate $\Vt_{m+1}^1$ and $\Ut_m^1$ from the radial vorticity 
equation (\ref{vort2}) and the divergence equation (\ref{div2}). 
In view of the previous results, this last equation takes the form of 
eqns. (\ref{vort1}) and (\ref{div1}) with the simple changes 
$\Ut_m^0 \to \Ut_m^1$ and $\Vt_{m+1}^0 \to \Vt_{m+1}^1$.
The  coefficients are (again) related by eqns.~(\ref{coeff1}) and (\ref{coeff3}).

The relevant boundary conditions are equations (\ref{axbc2})
and (\ref{pobc2}) at $r=R_c$ and $r=R$. Their combination leads to,
\bear
A_u &=& -\frac{(m-1) A\lambda_3^2 R^m}{2\mut^{1/2} \sinh(\lambda_1\Delta) 
\lambda_1 (\lambda_1^2 + \lambda_3^2)} \left [ 1 -\left (\frac{R_c}{R}\right )^m  
e^{-\lambda_1\Delta} \right ] 
\label{Au}
\\
\nonumber \\
B_u &=& -\frac{(m-1) A \lambda_3^2 R^m}{2\mut^{1/2}\sinh(\lambda_1\Delta)
\lambda_1 (\lambda_1^2 + \lambda_3^2)} \left [ 1 -\left (\frac{R_c}{R}\right )^m  
e^{\lambda_1\Delta} \right ]
\label{Bu}
\\
\nonumber \\
C_u &=& \frac{(m-1) A \lambda_1^2 R^m }{\mut^{1/2} \sin(\lambda_3 \Delta) 
\lambda_3 (\lambda_1^2 + \lambda_3^2) } 
\left [ 1 -\left (\frac{R_c}{R} \right )^m \cos(\lambda_3\Delta) \right ]
\label{Cu}
\\
\nonumber\\
D_u &=& \frac{(m-1) A \lambda_1^2}{\mut^{1/2} \lambda_3 (\lambda_1^2
+ \lambda_3^2 )} R_c^m
\label{Du}
\eear

Note that in the limit of a thick crust we have 
$e^{\lambda_1 \Delta} \gg 1 $, in which case we get $ A_u, A_v \to 0 $  
i.e. the growing exponentials are eliminated. Analogously, the two solutions that 
grow away from the surface will also vanish in this limit.

The remaining components $V_{m+1}^1$, $U_m^1$ and $W_{m+1}^1$ 
can be obtained from equations (\ref{cont3}), (\ref{vort3}) and (\ref{div3}).
Neglecting all homogeneous solutions, since we are only interested in the 
flow driven by the leading order r-mode, we find 

\bear
U_m^1 &=& 0
\label{U1} \\
V_{m+1}^1 &=& \frac{A \sigma_1}{2\Omega \mut^{1/2}} 
(m+1) Q_{m+1} r^{m+1}
\label{V1}
\\
\nonumber \\
W_{m+1}^1 &=& (m+1) V_{m+1}^1 
\label{W1}
\eear
For the pressure eigenfunctions we again find that $\htt_{m+1}^1 = 0$
identically (due to eqns. (\ref{vort2}), (\ref{div2}) and the vanishing 
leading order flow). On the other hand,
\be
h_{m+1}^1 = -\frac{2\Omega A Q_{m+1} \sigma_1 }{\mut^{1/2}}
\frac{(2m^2 + m-1)}{(m+1)^2}~r^{m+1}
\ee
    
The flows in the neutron star core and crust ``communicate''
via the interface conditions. In view of our previous results, some of these 
conditions reduce to trivial identities. The remaining non-trivial conditions 
are eqns.~(\ref{tract3}) and (\ref{tract4}) which fix $A$ and $\gamma$, 
\be
A=1, \qquad \gamma=0
\ee

The requirement for the radial traction to vanish on the surface
gives, at leading order,
\be
h_{m+1}^0 (R) = 0
\ee
This familiar condition {\it cannot} be satisfied by our leading
order solution for the simple reason that we have assumed a purely
axial r-mode background flow in the core. Indeed, in order to be fully consistent, 
one would need to introduce ${\cal O}(\Omega^2)$ polar corrections, as for example 
in \cite{LU}. These pieces are neglected in our calculation and consequently the 
above surface condition should not be taken into account.

At the next order we have,
 \be
h_{m+1}^1 = -\frac{g_0}{R} W_{m+1}^1
\ee
from which we obtain an expression for $\sigma_1$,
\be
4\Omega^2 \sigma_1 ( 2m^2 +m -1 )  = 
\frac{g_0}{R} \sigma_1 (m+1)^4
\label{cond1}
\ee
The only way to satisfy this condition is by making $\sigma_1 =0$. Hence, the leading 
order correction to the frequency appears at ${\cal O}(\mut)$ (or higher).
This result has the consequence that for the induced {\it core flow},
\be
V_{m+1}^1 = W_{m+1}^1 = U_m^1 = 0
\ee 

At this point we have used all available dynamical equations and boundary conditions 
at ${\cal O}(\mut^{1/2})$. Our results allow us to compute the crust-core slippage, 
a crucial quantity for the determination of the Ekman viscous dissipation rate associated 
with the presence of the crust \cite{LU}. Unfortunately, as we have just found, the crust-induced 
frequency correction is undetermined at this level. If this quantity is desired, 
one has to extend the calculation to the next order.


\subsubsection{Second-order crust flow}

We have already worked out the second order corrections to the core
flow. In this section we deal with the corresponding crust corrections. 
To begin with, from the continuity equation (\ref{cont4}) we immediately
obtain,
\be
\Wt_{m+1}^2 (r) = \frac{(m+1)(m+2)}{\mut^{1/2}} \left [ A_v I_1(r) + B_v I_2(r)
+ C_v I_3(r) - D_v I_4(r) \right ] \label{Wt2}
\label{Wt2_b}
\ee
where we have defined the integrals,
\bear
I_1(r) &=& \int_{R_c}^r \frac{dr^\prime}{r^\prime} e^{\lambda_1(r^\prime -R_c)},
\qquad 
I_2(r) = \int_{R_c}^r \frac{dr^\prime}{r^\prime} e^{-\lambda_1(r^\prime -R_c)}
\label{I1}
\\
\nonumber \\
I_3(r) &=& \int_{R_c}^r \frac{dr^\prime}{r^\prime} \cos[\lambda_3 (r^\prime -R_c)]
\qquad 
I_4(r) = \int_{R_c}^r \frac{dr^\prime}{r^\prime} \sin[\lambda_3 (r^\prime -R_c)]
\label{I2}
\eear
and we have used the continuity of the  radial displacement (\ref{Wbc3}) to fix the
integration constant such that $\Wt_{m+1}^2(R_c) = 0 $. 
Given this, condition (\ref{tract3}) requires that
\be
\htt_{m+1}^2(R_c) = 0
\label{cond3}
\ee

From eqns.~(\ref{cont5}), (\ref{vort4}) and (\ref{div5}) we get (again neglecting
all homogeneous solutions),
\bear
U_m^2 &=& 0
\\
\nonumber \\
V_{m+1}^2 &=& \frac{(m+1) Q_{m+1}}{2\Omega \mut} \sigma_2 ~r^{m+1}
\\
\nonumber \\
W_{m+1}^2 &=& (m+1) V_{m+1}^2
\eear

Moreover, from eqns.~(\ref{pres5}), (\ref{pres6}) in combination with
eqns.~(\ref{div4}), (\ref{cont4}) and (\ref{vort5}) we obtain the
pressure corrections,
\bear
\htt_{m+1}^2 &=& -\frac{4\Omega^2 m}{(m+1) \mut^{1/2}}  [ \{ Q_{m+1} A_u
+ A_v \} I_1(r)  + \{ Q_{m+1} B_u + B_v \} I_2(r) + \{ Q_{m+1} C_u + C_v
\}  I_3(r)
\label{htt_b}
\nonumber \\
\nonumber \\ 
&-& \{ Q_{m+1} D_u + D_v \} I_4(r) ] 
\\
\nonumber \\
h_{m+1}^2 &=& \frac{2\Omega Q_{m+1}}{\mut} \frac{2-m(2m +3)}
{(m+1)(m+2)}\sigma_2 r^{m+1}
\eear

The only remaining non-trivial condition (at this level of approximation) is
equation (\ref{tract6}) which can be solved to provide the frequency correction 
\be
\sigma_2 = -\frac{\mut R^{-(m+1)}}{2\Omega Q_{m+1}} \left [ \Omega_0^2 \Wt_{m+1}^2(R)
+ \htt_{m+1}^2(R) \right ] \left [ \frac{2-m(2m+3)}{(m+1)(m+2)}
+ \frac{1}{4} (m+1)^2 \left (\frac{\Omega_0}{\Omega} \right )^2  \right ]^{-1}
\label{sigma2}
\ee


\section{Final results}
\label{sec:results}

After a somewhat involved calculation, we are now in a position where we can summarise the 
results and extract physically interesting quantities. 
The two most relevant quantities are  i) the core-crust slippage,
which enters in the estimates of the Ekman layer damping rate of 
core r-mode oscillations, and ii) the crust-induced frequency correction.
The latter is interesting because it will unveil the presence of parameter
values where our approximation scheme breaks down. As we will discuss, 
this happens because of the
existence of distinct shear modes in the crust.  

\begin{figure}[ht] 
\centerline{\includegraphics[height=6cm,clip] {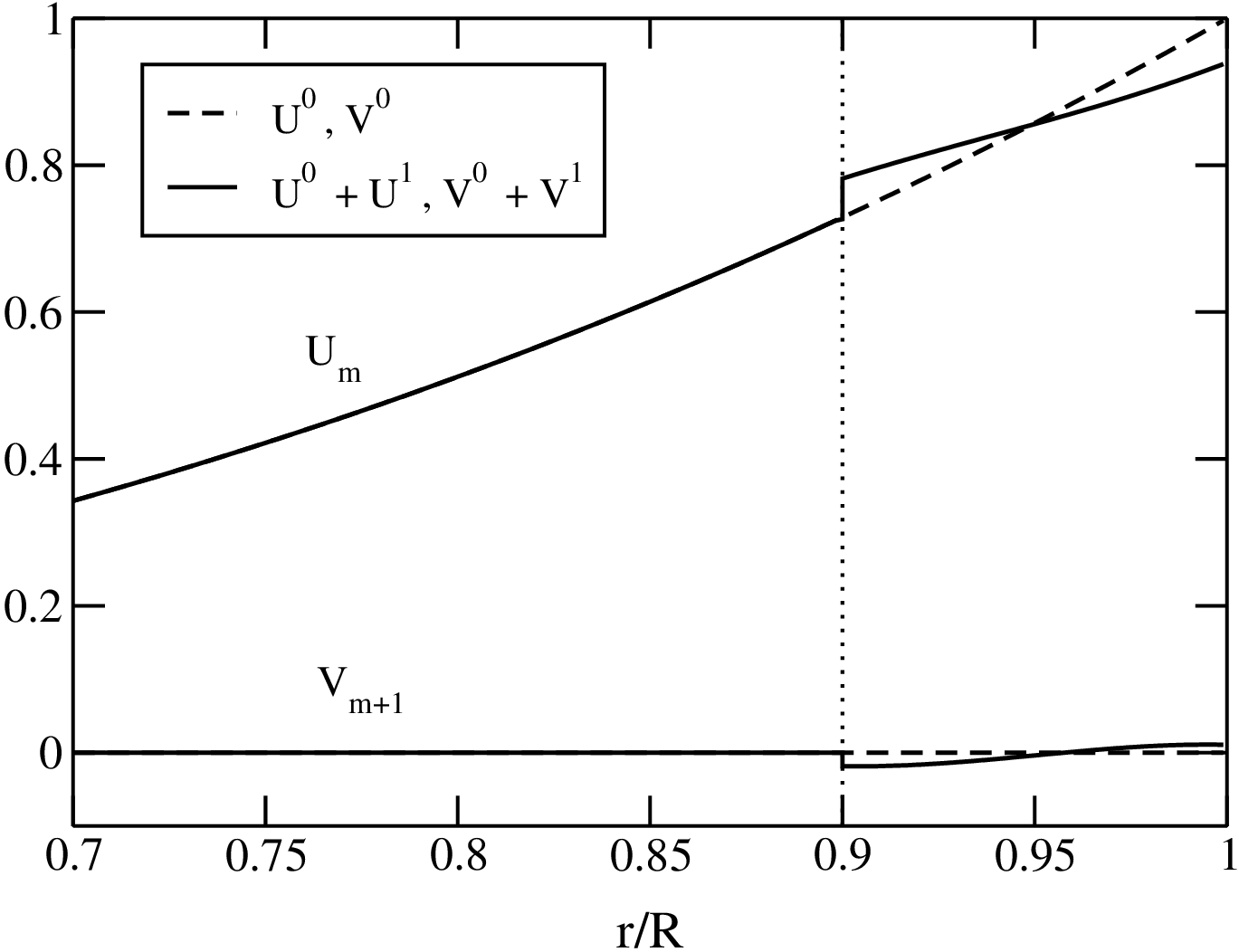} }
\caption{Eigenfunctions $U_m$ and $V_{m+1}$ as functions of $r/R$ 
for the $l=m=2$ r-mode and for $\Omega = 0.3 \Omega_0$, $\mut = 10^{-4} R^2 \Omega_0^2$. 
Dashed and solid curves represent the background and ${\cal O}(\mut^{1/2})$ 
corrected eigenfunctions, respectively. The vertical dotted line represents 
the crust-core boundary, here taken to be at $R_c =0.9 R$.} 
\label{fig1}  
\end{figure}  

We begin by considering the extent to which the crust moves along with the core fluid. 
From the studies of the Ekman layer problem (see \cite{Ekman} for references)
we know that the potential presence of 
a discontinuity at the core-crust interface could have significant impact on 
the viscous damping of a core oscillation. In principle, the role of viscosity is 
to smooth out any discontinuities in the inviscid solution. To date, there have been 
no detailed studies of the viscous problem for a star with an elastic crust. 
Instead the effect that the elasticity of the crust has on the Ekman 
damping timescale has been estimated by the core-crust ``slippage'' \cite{LU}.
The magnitude of the relative displacement $\Delta \xi$ between the core and 
the crust is (obviously) given by
\be
\Delta {\bf \xi} = {\bf \hat{e}_\theta} [\xi^\theta(R_c^{+}) -\xi^\theta(R_c^{-})  ]
 + {\bf \hat{e}_\varphi} [\xi^\varphi(R_c^{+}) -\xi^\varphi(R_c^{-})  ]
\ee
The relative contribution of the polar and axial pieces  
can be seen in Fig.~\ref{fig1} where we show the $m=2$ eigenfunctions for a fixed angular 
frequency $\Omega =0.3\Omega_0$ (about half the Kepler break up rate) and  
$\mut = 10^{-4} R^2 \Omega_0^2$. For this particular rotational frequency the axial
piece is dominant, although we note that the polar piece becomes dominant 
once we increase the rotation rate and cross the first of the crust-core 
resonances discussed later. It is evident from the results shown in Fig.~\ref{fig1} 
that the corrections to the basic $l=m $ r-mode oscillation are everywhere small,
thus providing credence to our approximation of truncating the multipole expansion.   

\begin{figure}[ht] 
\centerline{\includegraphics[height=6cm,clip] {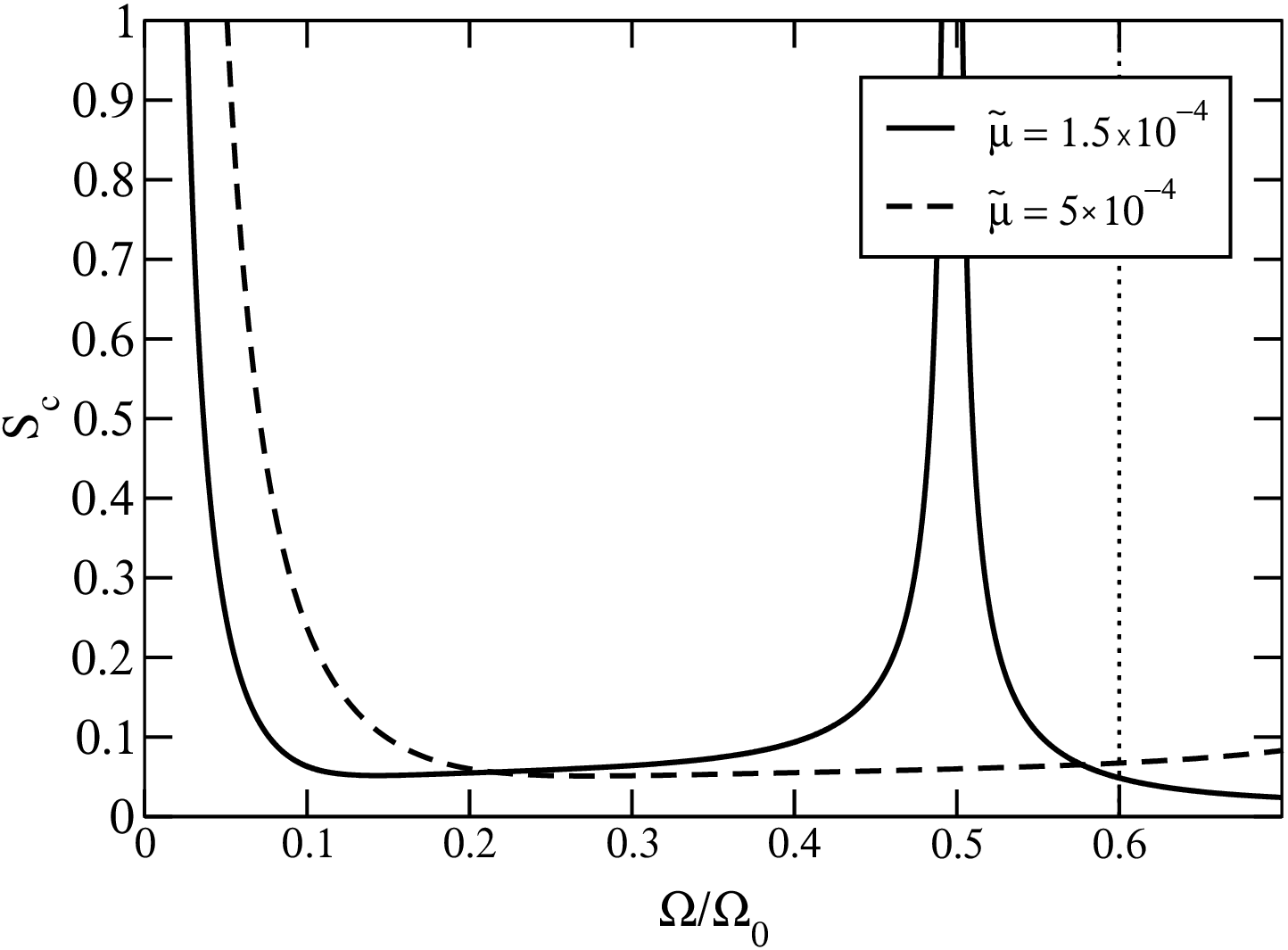} }
\centerline{\includegraphics[height=6cm,clip] {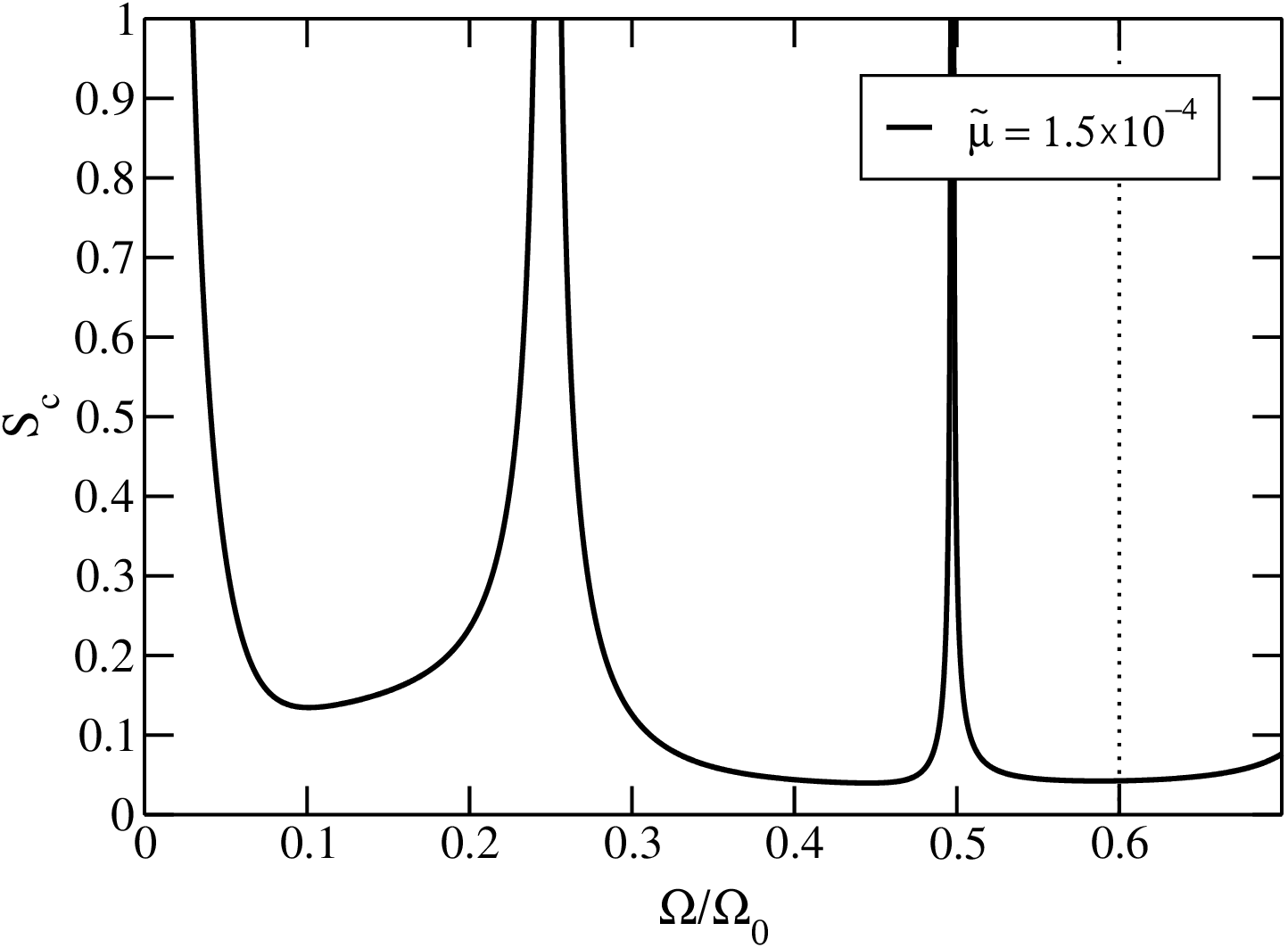} }
\caption{The slippage ${\cal S}_c$ as function of $\Omega/\Omega_0$ for 
$R_c = 0.9 R$ (top) and $ R_c =0.8R$ (bottom). Note that the shear modulus $\mut$
is expressed in units of $R^2 \Omega_0^2$. The vertical dotted
line marks the Kepler break-up limit $\Omega_K \approx 0.6 \Omega_0$. 
In both panels we have set $l=m=2$ for the core r-mode oscillation.}
\label{fig2}   
\end{figure}

The figure also clearly illustrates the discontinuity at  $r=R_c$. 
We  define the crust-core  slippage ${\cal S}_c$ by normalising the corresponding
relative displacement with respect to the standard r-mode displacement $\xi_0$. This leads to 
\be
{\cal S}_c \equiv   \left| \frac{ \Delta \xi}{\xi_0} \right| = 
\frac{1}{R_c^{m+1}} \left [\, \left| U_{m}(R_c^{+}) -U_{m}(R_c^{-}) \right|^2
+ \left| V_{m+1}(R_c^{+}) -V_{m+1}(R_c^{-}) \right|^2\, \right ]^{1/2}
\label{slip}
\ee
For a rigid crust the slippage is obviously unity, since
$ {\Delta \xi } = {\bf \xi}_0$ if the fluid satisfies a no-slip condition at the base of the crust. 
Within the framework of the present
calculation (a  crust with $ \Lambda \ll 1$), we can use
the results for the ${\cal O}(\mut^{1/2})$ flow (Sections~\ref{sec:crustflow} \& 
\ref{sec:coreflow}) and write the leading order expression for the slippage.    
We have found that the corrected global r-mode eigenfunctions are,
\bear
U_m(r) &=&  r^{m+1} + \Theta(\zeta)~\mut^{1/2} \left [ A_u e^{\lambda_1 \zeta} + 
B_u e^{-\lambda_1 \zeta} + C_u \cos(\lambda_3 \zeta) 
-D_u \sin(\lambda_3 \zeta)  \right ]
\\
\nonumber \\
V_{m+1}(r) &=& \Theta(\zeta)~\mut^{1/2} \left [ A_v e^{\lambda_1 \zeta} + 
B_v e^{-\lambda_1 \zeta} + C_v \cos(\lambda_3 \zeta) 
-D_v \sin(\lambda_3 \zeta) \right ]
\\
\nonumber \\
W_{m+1}(r) &=& 0 + {\cal O}(\mut)
\eear
where $\lambda_1 $ and $\lambda_3$ are defined by (\ref{root1}) and (\ref{root2}), and the 
coefficients $A_u$ etcetera follow from (\ref{Au})--(\ref{Du}) and (\ref{coeff1}),(\ref{coeff3}).
 $\Theta(\zeta)$ represents the unit step function and as before $\zeta =r-R_c$. 
From the above formulae we see that the discontinuity at the core-crust interface follows from
\bear
U_{m}(R_c^{+}) -U_{m}(R_c^{-}) &=& \mut^{1/2} \Ut_{m}^1(R_c)
= \mut^{1/2} ( A_u + B_u + C_u )
\\
\nonumber \\
V_{m+1}(R_c^{+}) -V_{m+1}(R_c^{-}) &=& \mut^{1/2} \Vt_{m+1}^1(R_c)
= \mut^{1/2} ( A_v + B_v + C_v )
\eear
Then (\ref{slip}) becomes,
\be
{\cal S}_c = \frac{\tilde{\mu}^{1/2}}{R_c^{m+1}} \left [\, 
\left| A_u + B_u + C_u \right|^2 + \left| A_v + B_v + C_v \right|^2\, \right ]^{1/2}
\label{slip2}
\ee
The way the slippage enters the r-mode problem has been discussed by LU \cite{LU} and Kinney \& Mendell
\cite{kinney}. They find that the slippage enters quadratically in the r-mode damping formula,
which leads to a revised Ekman layer timescale
\be
\tau_\mathrm{E} \to { \tau_\mathrm{E}  \over {\cal S}_c^2} 
\label{scale}
\ee 
In Fig.~\ref{fig2} we show the slippage as a function of $\Omega$, 
for $m=2$ and for two values of the crust thickness. The  results show that 
a typical value is ${\cal S}_c \approx 0.05$. In other words, we estimate that the
r-mode damping time is about a factor 400 {\it longer} in the case of an elastic crust 
that partakes in the mode oscillation than in the case of a solid crust.  
This estimate is in good agreement with that obtained by LU.

\begin{figure}[ht] 
\centerline{\includegraphics[height=6cm,clip] {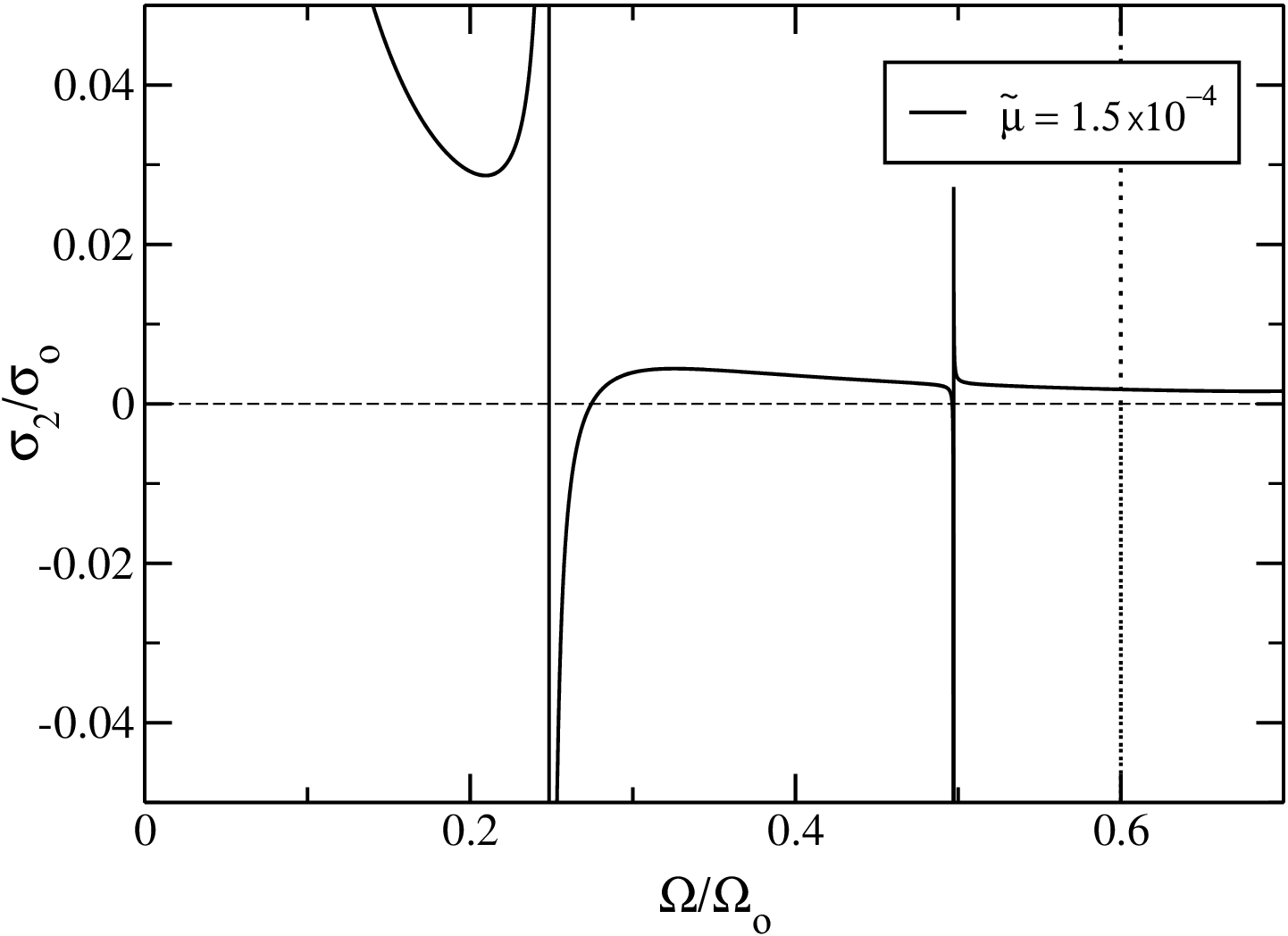} }
\caption{Mode frequency correction $\sigma_2 $ (normalised in units
of the leading order frequency $\sigma_0$) as function of $\Omega/\Omega_0$, 
for $l=m=2$ and $R_c=0.8R$. The displayed value for $\mut$ is in units of $R^2 \Omega_0^2$ and the 
vertical dotted line marks the Kepler break-up limit.} 
\label{fig3}  
\end{figure}

Fig.~\ref{fig2} also shows the presence of resonances at certain discrete values of the angular frequency. 
These peaks correspond to, cf. (\ref{reson}),  
\be
\lambda_3 (\Omega_n) \Delta = n \pi, \quad n \quad \mbox{integer}
\ee  
Solving this condition we find the resonant frequencies,
\be
\Omega_n = \mut^{1/2}\frac{n\pi}{2\Delta}(m+1)\sqrt{\frac{m+2}{\sqrt{\alpha}+1}} 
\label{reson2}  
\ee
These resonance points are apparent if we graph the frequency correction
$\sigma_2 $ from eqn.~(\ref{sigma2}), see Fig.~\ref{fig3}. 

Clearly, the number of resonant points  in the interval $\Omega \in [0,\Omega_0]$  
increases with increasing crust thickness and decreasing $\mut$. In the vicinity
of the resonances our calculation breaks down.
This means that, in the vicinity of these parameter values our 
starting assumption that the motion is dominated by the 
Coriolis force and can be represented as an r-mode plus a small correction is no longer valid. 
One would expect this to happen when the core r-mode frequency is 
close to the various crust mode frequencies. In other words, the resonances in 
Figs.~\ref{fig2} and \ref{fig3} indicate the occurrence of avoided crossings between 
core and crust modes. That this interpretation is correct  is clear from the results of LU 
\cite{LU} and Lee \& Yoshida \cite{yoshida}. Away from these points the mode frequency 
is essentially that of the core r-mode, in agreement with the numerical results of LU. Having said this, the careful reader may note that our small frequency correction differs in sign from that of LU for fast rotation. That the two results differ in this regime is not too surprising, and is likely due to LU including some higher order rotational corrections.

\section{Concluding remarks}
\label{sec:conclusions}

In this paper we have presented an analytic investigation into the interaction
between an r-mode core oscillation and an idealised elastic crust for a uniform density, 
incompressible, rotating neutron star. The essence of our calculation is the use of boundary 
layer theory techniques. Our results are in agreement with previous numerical treatments of the 
problem \cite{LU,yoshida}. They also offer a more detailed description of the workings of the crust-core coupling 
and the emergence of resonances between the r-mode and the various crustal modes, which leads
to avoided crossings in the numerical studies \cite{LU,yoshida}. We have provided simple analytic results 
for quantities like the crust-core slippage and resonant frequencies (eqns.~(\ref{slip2}) and (\ref{reson2})) 
which are of importance if one wants to estimate the viscous damping timescale for an unstable r-mode. 

There are several ways that one can refine and/or expand the present analysis. 
Recall that we have taken the fluid to be inviscid. This assumption could be relaxed and the 
resulting formulation should allow a consistent computation of the viscous dissipation rate. 
Note that presently this dissipation has only been estimated by rescaling the Ekman damping
rate with the slippage factor \cite{LU}. Other ways to improve our model would 
be to consider a crust with a non-uniform shear modulus (along the lines of \cite{LS}),
and/or a non-spherical rotating star. Both options would make an already complicated 
calculation even more complicated, and one would have to approach the problem numerically.      

An interesting extension would be the description of  a neutron star with an elastic 
crust {\it and} a magnetic field. Based on the results from the study of the magnetic Ekman problem 
\cite{mendell1,kinney} it is clear that a magnetic field may significantly change 
the associated r-mode damping timescale. The combination of an elastic crust and a magnetic field could well 
provide the most important damping mechanism for any unstable stellar modes. 
So far this combination has been investigated in \cite{kinney}, but not within a fully consistent framework.
The crustal elasticity was only taken into account in the form of a slippage factor rescaling
(as in (\ref{scale})) the magneto-viscous dissipation rate computed for a solid boundary.  
Moreover, the core oscillation is assumed to be that of a normal, non-magnetic r-mode, even in the presence of 
strong magnetic fields which would most likely change the mode structure. There are also difficult 
issues of principle since the interior magnetic field structure is essentially unknown for 
neutron stars.

One would also have to account for the potential presence of 
superconducting protons in the core. A closely related problem arises from the fact that the crust is 
expected to be penetrated by superfluid neutrons. So far there have been no attempts to account for
this feature, even though one would  suspect that it could have both 
quantitative and qualitative impact. 
 
From this list of outstanding problems it is clear that, after quite a lot of algebra, 
we have only  made  some small progress towards the understanding of an area where many serious challenges remain.


\appendix

\section{Core flow}
\label{app:core}

Using the expansion (\ref{core_expan}) in the basic equations (\ref{eq1}), (\ref{eq2}) and (\ref{continuity}),
when applied in the core (i.e $\mu =0 $), we obtain at first-order, 
\bear
(m+1) \sigma_1 U_{m}^{0} &=& 2\Omega Q_{m+1} [ W^{1}_{m+1} 
+ (m+2) V_{m+1}^1 ]
\label{vort_core}
\\
\nonumber \\
m(m+2) Q_{m+1} [r  U_{m}^{1 \prime} -(m+1) U_{m}^{1}] &=& 2 r V_{m+1}^{1\prime}
-mr W_{m+1}^{1\prime} -(m+2) W_{m+1}^{1} + m(m+1)(m+2)V^{1}_{m+1}
\label{div_core}
\\
\nonumber \\
(r W^{1}_{m+1})^\prime &=& (m+1)(m+2) V_{m+1}^{1}
\eear
The corresponding second-order equations are,
\bear
(m+1) \sigma_2 U_m^0 + (m+1) \sigma_1 U_m^1 &=& 2\Omega Q_{m+1}
[ W_{m+1}^2 + (m+2) V_{m+1}^2 ]
\\
\nonumber \\
m(m+2) Q_{m+1} [ r U_m^{2\prime} -(m+1) U_m^2 ] &=& 
\frac{\sigma_1}{2\Omega} (m+1)(m+2) [ r V_{m+1}^{1\prime} -W_{m+1}^1 ]
+ m(m+1)(m+2) V_{m+1}^2 
\nonumber \\
\nonumber \\
&+& 2r V_{m+1}^{2\prime} -mr W_{m+1}^{2\prime}
-(m+2) W_{m+1}^2  
\eear
From eqn.~(\ref{vort_core}) we get, after using the continuity equation 
(\ref{continuity}) and disregarding the homogeneous solution which becomes singular 
at the origin,
\be
W^{1}_{m+1} = \frac{\sigma_1}{2\Omega} (m+1)^2 Q_{m+1} r^{m+1}
\ee
Then eqn.~(\ref{continuity}) gives,
\be
V^{1}_{m+1} = \frac{\sigma_1}{2\Omega} (m+1) Q_{m+1} r^{m+1}
\ee
Solving the remaining first-order equation (\ref{div_core}) we get,
\be
U_{m}^{1} = \gamma r^{m+1}
\ee
where $\gamma$ is a constant (since the source term in (\ref{div_core}) vanishes identically 
this is the homogeneous solution).

Similarly we find for the second-order flow,
\bear
W_{m+1}^2 &=& \frac{(m+1)^2 Q_{m+1}}{2\Omega} [\sigma_2 + \gamma\sigma_1 ] r^{m+1}
\\
\nonumber \\
V_{m+1}^2 &=& \frac{(m+1)Q_{m+1}}{2\Omega} [ \sigma_2 +\gamma\sigma_1] r^{m+1}
\\
\nonumber \\
U_m^2 &=& \delta r^{m+1}
\eear
with $\delta$ another integration constant.


\section{Crust flow}
\label{app:crust}

In this Appendix we insert the expansion (\ref{uniform}) in the main dynamical equations
(\ref{vort_crust}), (\ref{div_crust}) and (\ref{cont_crust}). We group together terms of
the same order in $\mut$ and then split those groups into ``boundary layer''
and background pieces. For clarity, we will discuss this process in detail only for the continuity equation 
(which is the simplest of our equations). For the remaining dynamical equations we just list the results.

Inserting the uniform expansion in (\ref{cont_crust}) we get
\bear
&& r[\Wt_{m+1}^{0\prime} + \mut^a \{ \Wt_{m+1}^{1\prime} + W_{m+1}^{1\prime}  \} 
+ \mut^{2a} \{ W_{m+1}^{2\prime} + \Wt_{m+1}^{2 \prime} \} 
+\mut^{3a} \Wt_{m+1}^{3\prime}  ] = -\Wt_{m+1}^0 
- \mut^a \{ W_{m+1}^1 + \Wt_{m+1}^1  \} 
\nonumber \\
\nonumber \\
&&-\mut^{2a} \{ W_{m+1}^2 + \Wt_{m+1}^2  \} 
+ (m+1)(m+2) [ \Vt_{m+1}^0 + \mut^a \{ V_{m+1}^1 + \Vt_{m+1}^1  \} 
+ \mut^{2a} \{ V_{m+1}^2 + \Vt_{m+1}^2  \} ] 
+ {\cal O}(\mut^{3a})
\eear
Note the appearance of the seemingly ``higher-order'' term $\mut^{3a} 
\Wt_{m+1}^{3 \prime}$. This term is, in fact,  of order $\mut^{2a}$ and therefore has 
to be included. Terms of this kind appear in other equations and 
the boundary conditions and one has to be aware (and cautious!) of their presence. 
The continuity equation splits into, 
\bear
\Wt_{m+1}^{0 \prime} &=& 0 ,\ 
\quad \mbox{at ${\cal O}(\mut^{-a})$ }
\label{cont1}
\\
\nonumber \\
r \mut^a \Wt_{m+1}^{1\prime} + \Wt_{m+1}^{0} &=& (m+1)(m+2) \Vt_{m+1}^{0}
,\ \quad \mbox{at $ {\cal O}(1)$ }
\label{cont2}
\\
\nonumber \\
r W_{m+1}^{1\prime} + W_{m+1}^1 + \Wt_{m+1}^1 &=& (m+1)(m+2) 
( V_{m+1}^1 + \Vt_{m+1}^1 ) -r\mut^a \Wt_{m+1}^{2\prime} ,\ 
\quad \mbox{at ${\cal O}(\mut^a)$ }
\label{cont2b}
\eear
The first of these equations immediately leads to
\be
\Wt_{m+1}^{0} =0 
\ee
As (\ref{cont2b}) contains both ``boundary layer'' and background variables, 
 we  split it further into the two equations
\bear
&& r W_{m+1}^{1\prime} + W_{m+1}^1  = (m+1)(m+2) V_{m+1}^1 
\label{cont3}
\\
\nonumber \\
&& r \mut^a \Wt_{m+1}^{2\prime} = (m+1)(m+2) \Vt_{m+1}^1
\label{cont4}                              
\eear  
The  same procedure will be applied in several equations to follow. Note 
that it should {\it not} be applied in the case of boundary conditions, 
where one \underline{must} keep  ``boundary layer'' and core variables together. 
At ${\cal O}(\mut^{2a}) $  the resulting background continuity equation is,
\be
r W_{m+1}^{2\prime} + W_{m+1}^2 = (m+1)(m+2) V_{m+1}^2
\label{cont5}
\ee
The respective ``boundary layer'' equation will not be needed in our calculation.  

Analogously we find that the radial vorticity equation implies (\ref{rfreq}), with $\sigma \to \sigma_0$,  
at leading order.
To derive the next order equation (which should govern the leading 
order flow) we need to determine the exponent $a$. 
Potential terms for this equation are,
\bear
&& \mut (m+1) \Ut_{m}^{0\prime\prime} + \left \{ \sigma_0 \sigma_1 (m+1)
+ \frac{1}{r^2} \mut m(m+1)^2 \right \} ( U_{m}^{0} + \Ut_{m}^{0})
\nonumber \\
\nonumber \\
&&- 2 \sigma_0 \Omega Q_{m+1} (m+2) \Vt_{m+1}^0 +(m+1) \mut^{1+a} 
\Ut_m^{1\prime\prime} =0 
\eear
After some experimentation we find that $a=1/2$ is the appropriate choice 
(other possibilities do not lead to consistent solutions). 
Then we get at $ {\cal O}(1)$; 
\be
-\mut (m+1) \Ut_{m}^{0\prime\prime} + 2\Omega \sigma_0  Q_{m+1} (m+2) 
\Vt_{m+1}^{0} =0 
\label{vort1} 
\ee
while at ${\cal O}(\mut^{1/2})$ we have
\bear
 -\mut(m+1) \Ut_m^{1} + 2\Omega\sigma_0 Q_{m+1}[ (m+2) \Vt_{m+1}^1
+ \Wt_{m+1}^{1} ] &=& (m+1)\mut^{-1/2}\sigma_0 \sigma_1 \Ut_{m}^0
\nonumber \\
&& -2\Omega \sigma_1 \mut^{-1/2} (m+2) Q_{m+1} \Vt_{m+1}^0
\label{vort2} 
\\
\nonumber \\
2\Omega \mut^{1/2} Q_{m+1} [ W_{m+1}^1 + (m+2) V_{m+1}^1 ] &=& (m+1) \sigma_1 U_m^0
\label{vort3}
\eear
One level up, at ${\cal O}(\mut)$, we have
\bear
&& \mut^{-1/2} (m+1) (\sigma_1^2 + \sigma_0 \sigma_2 ) U_m^0 + 
(m+1) \sigma_1\sigma_0 U_m^1  = -\mut^{1/2} (m+1) \left [ U_m^{0\prime\prime} 
-\frac{1}{r^2}m(m+1) U_m^0 \right ] 
\nonumber \\
\nonumber \\
&& + 2\Omega \sigma_1 Q_{m+1} 
[ W_{m+1}^1 + (m+2) V_{m+1}^1 ] 
+ 2\Omega \sigma_0 Q_{m+1} \mut^{1/2} 
[ W_{m+1}^2 + (m+2) V_{m+1}^2 ] 
\label{vort4}
\eear
and
\be
\mut^{3/2} (m+1) \Ut_{m}^{2\prime \prime} = 2\Omega \sigma_0 Q_{m+1}\mut^{1/2} 
[\Wt_{m+1}^2 + (m+2) \Vt_{m+1}^2 ] + 2\Omega \sigma_1 Q_{m+1}(m+2) \Vt_{m+1}^1 
-(m+1) \sigma_0 \sigma_1 \Ut_m^{1}
\label{vort5}
\ee

In the case of the ``divergence'' equation the leading order background flow is governed by the 
standard result (\ref{req2}), with $U_m \to U_m^0$. The ${\cal O}(\mut^{-1/2}) $,  
${\cal O}(\mut^0) $ and ${\cal O}(\mut^{1/2})$ equations are,
\be
(m+1)(m+2) [ \mut \Vt_{m+1}^{0\prime\prime\prime} + \sigma_0^2 \Vt_{m+1}^{0\prime} ] 
-2\Omega \sigma_0 m \Vt_{m+1}^{0\prime} -2\Omega\sigma_0  m(m+2) Q_{m+1} 
\Ut_m^{0\prime}=0
\label{div1}
\ee
\bear
&& (m+1)(m+2) [ r\mut^{3/2} \Vt_{m+1}^{1\prime\prime\prime} + \sigma_{0}^2 r\mut^{1/2} 
\Vt_{m+1}^{1\prime} + 2\sigma_0\sigma_1 r \Vt_{m+1}^{0\prime} 
+ 2m\Omega \sigma_0 \Vt_{m+1}^0 ] -2\Omega \sigma_0 mr \mut^{1/2} 
[ \Vt_{m+1}^{1\prime} + \Wt_{m+1}^{1\prime} ] 
\nonumber \\
\nonumber \\
&& -2\Omega \sigma_1 mr \Vt_{m+1}^{0\prime} -2\Omega \sigma_0 m(m+2) Q_{m+1} 
[ r\mut^{1/2} \Ut_{m}^{1\prime} -(m+1) \Ut_{m}^0 ] -2\Omega \sigma_1 m(m+2) Q_{m+1} 
r \Ut_{m}^{0\prime} =0
\label{div2}
\eear
and
\bear
&& (m+1)(m+2)\sigma_0^2 [r V_{m+1}^{1\prime} -W_{m+1}^1 ] + 2\Omega\sigma_0 
m(m+1)(m+2) V_{m+1}^1 = 2\Omega \sigma_0 mr [V_{m+1}^{1\prime} + W_{m+1}^{1\prime} ]
\nonumber \\
\nonumber \\
&& + 2\Omega \sigma_0 m(m+2) Q_{m+1} [ r U_m^{1\prime} -(m+1) U_m^1 ] 
+ 2\Omega \sigma_1 \mut^{-1/2} m (m+2) Q_{m+1} [ r U_m^{0\prime} -(m+1) U_m^0 ]
\label{div3}
\eear
\bear
&& (m+1)(m+2) r [ \mut^{3/2} \Vt_{m+1}^{2\prime\prime\prime} + \mut^{1/2} \sigma_0^2 
\Vt_{m+1}^{2\prime} ]
-2\Omega m\sigma_0 \mut^{1/2} r [ \Vt_{m+1}^{2\prime} + \Wt_{m+1}^{2\prime} ] =
2\Omega m(m+2) \sigma_0 Q_{m+1} \mut^{1/2} r \Ut_{m}^{2\prime}
\nonumber \\
\nonumber \\
&& + 2\Omega \sigma_1 mr \Vt_{m+1}^{1\prime} + 2\Omega m(m+2) Q_{m+1} 
[ \sigma_1 r \Ut_m^{1\prime} -\sigma_0 (m+1) \Ut_m^{1} ] 
-2\Omega m(m+1)(m+2) \sigma_0 \Vt_{m+1}^1
\label{div4}
\eear
respectively.
Finally, the ${\cal O}(\mut)$ equation is,
\bear 
&&m(m+2) Q_{m+1}[ r U_m^{2\prime} -(m+1) U_m^2 ] =
\frac{\sigma_1}{2\Omega} (m+1) (m+2) [ r V_{m+1}^{1\prime} - W_{m+1}^1 ]
+ 2r V_{m+1}^{2\prime}
\nonumber \\
\nonumber \\
&& + m(m+1)(m+2) V_{m+1}^2 -[mr W_{m+1}^{2\prime} +(m+2) W_{m+1}^2]
\label{div5}
\eear

To complete the required set of equations, we consider the
 $\theta$-component of the Euler equation which leads to the pressure eigenfunction
(\ref{pres1}) at  leading order. Meanwhile the ${\cal O}(\mut^0)$ and $ {\cal O}(\mut^{1/2})$ 
equations give
\bear
\htt_{m+1}^0 &=& \frac{1}{(m+2) Q_{m+1}} [ -m\sigma_0 \{ \sigma_0 -2\Omega Q_{m+1}^2  
\} \Ut_m^0 -m\mut \Ut_{m}^{0\prime\prime} + \sigma_0 Q_{m+1} \{ (m+2)\sigma_0 
+2m\Omega \} 
\Vt_{m+1}^0
\nonumber \\ 
&+& \mut (m+2) Q_{m+1} \Vt_{m+1}^{0\prime\prime} ]
\label{pres2}
\eear
and
\bear
h_{m+1}^1 &=& \frac{1}{(m+2) Q_{m+1}} [ -2m \sigma_1 \mut^{-1/2} \{ \sigma_0 - 
\Omega Q^2_{m+1} \} U_m^0 -m\sigma_0 \{ \sigma_0 -2\Omega Q^2_{m+1} \} U_{m}^1
\nonumber \\
\nonumber \\
 &+&  \sigma_0 Q_{m+1} \{ \sigma_0 (m+2) + 2m\Omega  \} V_{m+1}^1 ]
\label{pres3}
\\
\nonumber \\
\nonumber \\
\htt_{m+1}^1 &=& \frac{1}{(m+2) Q_{m+1}} [ -m\mut \Ut_{m}^{1\prime\prime} + 
(m+2) Q_{m+1} \mut \Vt_{m+1}^{1 \prime\prime} -m\sigma_0 \{ \sigma_0 -2\Omega 
Q_{m+1}^2 \}
\Ut_{m}^1 
\nonumber \\
\nonumber \\
 &+&  \sigma_0 Q_{m+1} \{ \sigma_0 (m+2) + 2m\Omega  \} \Vt_{m+1}^1  
- 2m\sigma_1 \mut^{-1/2} \{ \sigma_0 -\Omega Q_{m+1}^2 \} \Ut_{m}^0
\nonumber \\
\nonumber \\
&+& 2\sigma_1 \mut^{-1/2} Q_{m+1} \{ (m+2)\sigma_0 + m\Omega \} \Vt_{m+1}^0 ] 
\label{pres4}
\eear
The ${\cal O}(\mut)$ equations are,
\bear
h_{m+1}^2 &=& \frac{1}{(m+2) Q_{m+1}} \left [ -m\sigma_0 \{ \sigma_0 
-2\Omega Q^2_{m+1}  \} U_m^2 -\frac{2m\sigma_1}{\mut^{1/2}} \{ \sigma_0 
-\Omega Q_{m+1}^2 \} U_m^1 -\frac{m}{\mut} \{ \sigma_1^2 + 2\sigma_0\sigma_2 
-2\Omega \sigma_2 Q_{m+1}^2 \} U_m^0 \right.  
\nonumber \\
\nonumber \\
&+& \left. \sigma_0 Q_{m+1} \{ \sigma_0 (m+2) + 2m\Omega  \} V_{m+1}^2
+ \frac{2\sigma_1}{\mut^{1/2}} Q_{m+1} \{ \sigma_0 (m+2) + 2m\Omega  \} V_{m+1}^1  
\right ]
\label{pres5}
\eear
and
\bear
\htt_{m+1}^2 &=& \frac{1}{(m+2) Q_{m+1}} \left [  -m\sigma_0 \{ \sigma_0 
-2\Omega Q^2_{m+1}  \} \Ut_m^2 -\frac{2m\sigma_1}{\mut^{1/2}} \{ \sigma_0 
-\Omega Q_{m+1}^2 \} \Ut_m^1 + \sigma_0 Q_{m+1} \{ \sigma_0 (m+2) + 2m\Omega  \} 
\Vt_{m+1}^2 \right.
\nonumber \\
\nonumber \\
&& \left.  + \frac{2\sigma_1}{\mut^{1/2}} Q_{m+1} \{ \sigma_0 (m+2) + 2m\Omega  
\} \Vt_{m+1}^1 -m\mut \Ut_m^{2\prime\prime}+ \mut (m+2) Q_{m+1} 
\Vt_{m+1}^{2\prime\prime}
\right ]
\label{pres6}
\eear


\section{Boundary conditions}
\label{app:bcs}

In this Appendix we expand the boundary conditions (\ref{bc1})--(\ref{bc5}) with respect
to $\mut$.

From the conditions pertaining to the tangential traction at both $r=R_c$ and $r=R$ we obtain, 
\bear
\Ut_m^{0\prime} &=&0   \quad ,\ \mbox{at ${\cal O}(\mut^{-1/2})$   }
\label{axbc1}
\\
\nonumber \\
r [ U_m^{0\prime} + \mut^{1/2} \Ut_m^{1\prime}] &=& 2 ( U_m^0 + \Ut_m^0 ) 
\quad ,\ \mbox{at ${\cal O}(\mut^0)$  }
\label{axbc2}
\\
\nonumber \\
r [ U_m^{1\prime} + \mut^{1/2} \Ut_m^{2\prime} ] &=&  2 ( U_m^1 + \Ut_m^1 )  
\quad ,\ \mbox{ at $ {\cal O}(\mut^{1/2})$  }
\label{axbc3}
\\
\nonumber \\
\Vt_{m+1}^{0\prime} &=& 0 \quad ,\ \mbox{at $ {\cal O}(\mut^{-1/2})$ }
\label{pobc1}
\\
\nonumber \\
r \mut^{1/2} \Vt_{m+1}^{1\prime} &=& 2\Vt_{m+1}^0 \quad ,\ 
\mbox{at ${\cal O}(\mut^0)$ }
\label{pobc2}
\\
\nonumber \\ 
r [ V_{m+1}^{1\prime} +\mut^{1/2} \Vt_{m+1}^{2\prime} ]  &=& 2 (V_{m+1}^1 
+ \Vt_{m+1}^1 ) -( W_{m+1}^1 + \Wt_{m+1}^1 )  \quad ,\ 
\mbox{at ${\cal O}(\mut^{1/2}) $ } 
\label{pobc3}
\eear
 
Continuity of the radial displacement at $r= R_c$ demands,
\bear
W_{m+1}^{0}(R_c^{-}) &=& W_{m+1}^{0}(R_c^{+}) + \Wt_{m+1}^{0}(R_c^{+})
\label{Wbc1} 
\\
W_{m+1}^1 (R_c^{-}) &=&  \mut^{1/2} [W_{m+1}^{1}(R_c^{+}) 
+ \Wt_{m+1}^{1}(R_c{+}) ]
\label{Wbc2}
\\
W_{m+1}^2 (R_c^{-}) &=&  \mut  [ W_{m+1}^{2}(R_c^{+}) 
+ \Wt_{m+1}^{2}(R_c^{+})] 
\label{Wbc3} 
\eear
Note that the first of these conditions is trivially satisfied since we are assuming 
that the leading order flow is purely axial.

Finally, from the radial traction conditions at the crust-core interface
and the surface we get, 
\bear
h_{m+1}^{0}(R_c^{-}) &=& h_{m+1}^{0}(R_c^{+}) + \htt_{m+1}^{0}(R_c^{+}) 
\label{tract1}
\\
h_{m+1}^{1}(R_c^{-}) &=& \mut^{1/2} [ h_{m+1}^1 + \htt_{m+1}^1  ](R_c^{+})
\label{tract2}
\\
h_{m+1}^{2}(R_c^{-}) &=& \mut [ h_{m+1}^2 + \htt_{m+1}^2 ](R_c^{+})
-2\mut^{3/2} \frac{\Wt_{m+1}^{1\prime}(R_c^{+})}{R_c^{+}}
\label{tract3}
\eear
,and
\bear
h_{m+1}^{0} + \htt_{m+1}^0  &=& 0
\label{tract4}
\\
h_{m+1}^{1} + \htt_{m+1}^{1} &=& -\frac{g_0}{R} [ W_{m+1}^{1}
+ \Wt_{m+1}^1 ] 
\label{tract5}
\\
h^{2}_{m+1} + \htt_{m+1}^2 &=& -\frac{g_0}{R} [ W_{m+1}^2 + \Wt_{m+1}^2 ]
+ \frac{2\mut^{1/2}}{R} \Wt_{m+1}^{1\prime}
\label{tract6}
\eear
respectively.


\acknowledgments

This work was supported by PPARC through grant number PPA/G/S/2002/00038.
NA also acknowledges support from PPARC via Senior Research Fellowship no
PP/C505791/1.




\begin{thebibliography}{99}

\bibitem{na98}
N. Andersson, ApJ {\bf 502}, 708 (1998)

\bibitem{morsink}
J.L. Friedman and S.M. Morsink, ApJ {\bf 502}, 714 (1998)

\bibitem{na_rev}
N. Andersson and K.D. Kokkotas, Int. J. Mod. Phys. {\bf D10}, 381 (2001)

\bibitem{Ekman}
K. Glampedakis and N. Andersson, MNRAS {\bf 371}, 1311 (2006)

\bibitem{haensel}
 P. Haensel, K.P. Levenfish  and  D.G. Yakovlev, Astron. Astrophys. {\bf 381}, 1080 (2002)

\bibitem{reis}
A. Reisenegger and A. Bonacic, Phys. Rev. Lett {\bf 91},  201103 (2003)

\bibitem{owen2}
M. Nayyar and  B.J. Owen, Phys. Rev. D {\bf 73}, 084001 (2006)

\bibitem{shear} 
N. Andersson,  G.L. Comer and K. Glampedakis, Nucl. Phys. A {\bf 763},  212 (2005)
 

\bibitem{LS}
U. Lee and T.E. Strohmayer, Astron. Astrophys.  {\bf 311}, 155 (1996)

\bibitem{mcdermott}
P.N. McDermott,  H.M. Van Horn and  C.J. Hansen, ApJ {\bf 325}, 725 (1988)

\bibitem{strohmayer}
 T.E. Strohmayer, ApJ, {\bf 372}, 573 (1991)

\bibitem{yoshida}
S. Yoshida and U. Lee, ApJ {\bf 546}, 1121 (2001)

\bibitem{LU}
 Y. Levin and G. Ushomirsky, MNRAS, {\bf 324}, 917 (2001) 

\bibitem{douche} 
F. Douchin and P. Haensel, Astron. Astrophys. {\bf 380}, 151 (2001)


\bibitem{lockitch}
K.H. Lockitch and J.L. Friedman, ApJ {\bf 521}, 764 (1999)


\bibitem{landau}
L.D. Landau and  E.M. Lifshitz, 
{\it Theory of Elasticity}, Course of Theoretical 
Physics Vol. 7, 3rd edition (Butterworth, Heinemann,
Oxford 1999). 


\bibitem{orszag}
 C.M. Bender and  S.A. Orszag, {\it Advanced Mathematical Methods for Scientists
and Engineers} (McGraw-Hill International Editions, 1978)
 

\bibitem{kinney}
J. Kinney and G. Mendell, Phys. Rev. D {\bf 67} 024032 (2003)

\bibitem{mendell1}
G. Mendell, Phys. Rev. D {\bf 64}, 044009 (2001)

\end{thebibliography}
\end{document}